\documentclass[twocolumn,aps,superscriptaddress,amsmath,showpacs,prl,floatfix]{revtex4-1}
\usepackage{graphicx}
\usepackage{amssymb}
\usepackage{hyperref}
\usepackage{amsmath}
\usepackage{dcolumn}
\usepackage{comment}
\usepackage{color}

\begin{document}

\sloppy 

\title{Polymer adsorption on vicinal surfaces}
\author{Nathann T. Rodrigues}
\email{nathann.rodrigues@unb.br}
\affiliation{International Center of Physics, Institute of Physics, University of Brasilia, 70910-900, Brasilia, Federal District, Brazil} 
\author{Tiago J. Oliveira}
\email{tiago@ufv.br}
\affiliation{Departamento de F\'isica, Universidade Federal de Vi\c cosa, 36570-900, Viçosa, Minas Gerais, Brazil}

\date{November 11, 2026}

\begin{abstract}
While polymer adsorption is theoretically well established on flat surfaces --- and somewhat understood on rough surfaces --- the common and practically relevant case of nearly flat stepped surfaces remains poorly studied. Here, we address this gap via extensive simulations of flexible polymer chains grafted at the step edge of vicinal surfaces. For wide terraces, the polymer-step interaction yields a stable step-attached (SA) phase, alongside the desorbed and terrace-adsorbed (TA) phases, all of them separated by continuous transition lines that meet at a multicritical point previously unanticipated in polymer thermodynamics. This scenario is observed for terrace widths $\mathcal{W} \gtrsim 3 R_g^{(2D)}$, being $R_g^{(2D)}$ the radius of gyration of TA polymers. Simpler phase diagrams, absent this multicriticality, are found for $\mathcal{W} \lesssim 3 R_g^{(2D)}$, where a simple crossover exists between the SA and TA phases. The presence of kinks at the steps does not alter these behaviors. Besides their theoretical relevance, these findings unveil an interesting route to produce polymer nanowires and control polymer conformation via step attachment-detachment on designed stepped surfaces.
\end{abstract}


\maketitle

{\bf\textit{Introduction.}} Polymer adsorption at surfaces is key to a vast number of industrial and biological processes, including adhesion \cite{Adhesion}, lubrication \cite{Lubrication}, stabilization of colloidal suspensions \cite{StabColloids}, fabrication of biosensors \cite{sensors}, drug delivery systems \cite{drugdeliv}, and many others \cite{Applic}. This has motivated a large body of literature aimed at understanding the underlying physics of polymer adsorption. It is well known that when a polymer is exposed to an attractive surface, the competition between the energetic gain of attachment and the entropic loss of surface proximity drives the system through transitions from desorbed to adsorbed phases, depending on the polymer-substrate interaction strength ($\varepsilon$) and the temperature ($T$) \cite{BachmannBook,AdsorpBook1,AdsorpBook2}. The great majority of these studies consider adsorption on perfectly flat and homogeneous surfaces \cite{BachmannBook,AdsorpBook1,AdsorpBook2}; yet, even this ideal situation remains a topic of recent debate regarding the critical properties at the ordinary \cite{Plascak,Chris1,Chris2,NT2019_1} and special adsorption transitions \cite{NT2019_1,NT2023}.

Surface inhomogeneities may also play a pivotal role in the adsorption process. This has been demonstrated, for instance, in several works theoretically analyzing polymer adsorption on perfectly flat but \textit{chemically inhomogeneous} surfaces --- formed by different functional groups arranged in checkerboard \cite{checkerboard}, stripe \cite{stripe}, and other regular patterns \cite{Muthukumar,Qi,Bachmann2}, as well as in random configurations \cite{randPattern} --- which are relevant for molecular recognition \cite{MolRec}.

Though similarly important, the effects of \textit{physical inhomogeneities} (arising from inexorable surface imperfections) on adsorption are not as well explored. Indeed, most existing theoretical works focus on locally smooth but globally rough planar surfaces featuring, for example, sinusoidal \cite{sinusSubst,Ji} or checkerboard \cite{checkbSubst} corrugations, or square-well holes \cite{squareHoles}. The same is true for polymer adsorption on non-planar geometries, such as the \textit{smooth} surfaces of spheres \cite{SupEsferica,Ji}, cylinders and rods \cite{SupCilind,BachmannBook}, mesoscopic particles \cite{MesoPart}, and repulsive wedges with an attractive edge \cite{Kardar}. Adsorption on planar substrates with disordered and fractal shapes has also been analyzed in a few studies \cite{disordSubst}.

At a microscopic level, solid surfaces are rarely atomically flat, nor do they always exhibit a large roughness. Often, the closest approximation to a flat system is a low-index facet featuring broad flat terraces separated by long steps --- so-called vicinal surfaces \cite{KrugBook}. Arrangements of steps and terraces are indeed ubiquitous in crystalline surfaces \cite{KrugBook,JimRev}. While the role of such steps in polymerization has been discussed in the literature \cite{StepPolymerization}, their effect on polymer adsorption has remained largely unappreciated, except for very short chains \cite{StepAdsorption}. We remark that when adatoms (atoms or small molecules) adsorb onto such surfaces and subsequently diffuse, the steps are critical to their aggregation process; among other reasons \cite{noteES}, this is because adatoms interact with the step in addition to the adatom-terrace interaction. Naturally, this step-induced effect should be negligible for long polymers if their persistence length, $\ell$, is much larger than the step height, $\mathcal{H}$. However, in many realistic scenarios, one may find $\ell \sim \mathcal{H}$, even for single atomic steps, where $\mathcal{H} \approx 0.2\text{--}0.3~\text{nm}$ for metallic and semiconductor surfaces \cite{Kittel}. Relevant examples of polymers with such a small $\ell$ include polyethylene glycol (PEG, $\ell \approx 0.35~\text{nm}$) \cite{Oesterhelt1999}, unfolded or intrinsically disordered proteins ($\ell \approx 0.5~\text{nm}$) \cite{Dietz2004}, polydimethylsiloxane (PDMS, $\ell \approx 0.5\text{--}0.7~\text{nm}$) \cite{Rubinstein2003}, and single-stranded DNA ($\ell \approx 0.75~\text{nm}$) \cite{Smith1996}. Moreover, $\mathcal{H}$ can be much larger than this due to step bunching or on microfabricated surfaces. Therefore, steps can dictate the macromolecular conformation at the surface for a broad class of polymers, offering an avenue to control their properties via the substrate.

In this Letter, we tackle this fundamental issue through a detailed numerical study of the adsorption of flexible linear homopolymer chains onto vicinal surfaces under dilute, good-solvent conditions. Analogous to adsorbed adatoms, the polymer monomers at the surface interact attractively  with both the terraces and the steps. We investigate vicinal surfaces with straight and kinked steps, for several terrace widths. Importantly, in addition to the well-known desorbed and terrace-adsorbed phases, we identify a stable step-attached phase and a previously unobserved desorbed-to-step-attached continuous phase transition, with unexpectedly large critical exponents. The phase diagrams of the system are fully characterized, revealing distinct behaviors for short versus wide terraces, and exposing a novel multicritical behavior for polymer adsorption in the latter case.


\begin{figure}[t]
\includegraphics[width=8.0cm]{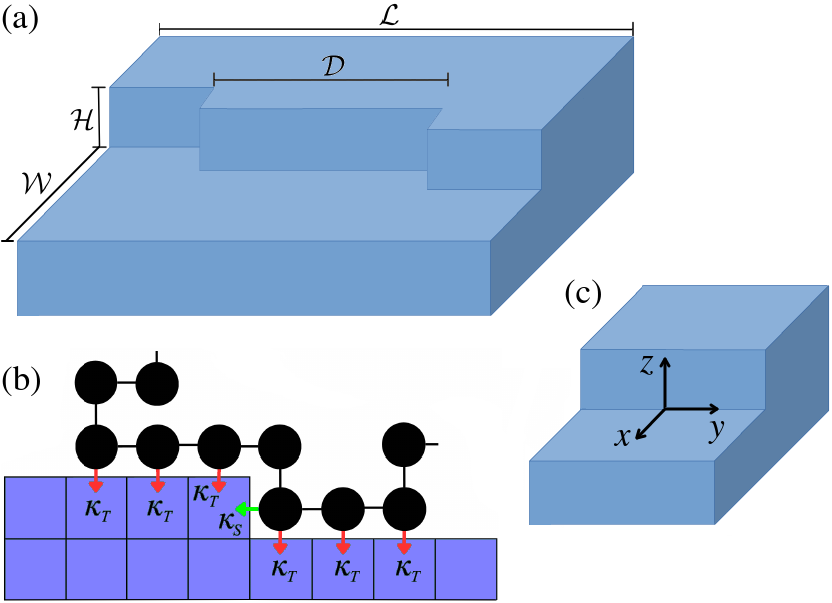}
\caption{(a) Schematic representation of a vicinal surface with terrace width $\mathcal{W}$ and steps of height $\mathcal{H}$, length $\mathcal{L}$ and distance between kinks $\mathcal{D}$. (b) Lateral view of a polymer near a step. Each monomer (black circle) contact with a terrace site is indicated by a red arrow, while the green arrow represent a contact with a step. (c) Definition of the coordinate system.}
\label{fig1}
\end{figure}

{\bf \textit{Model and Methods.}} As is standard in the literature \cite{Carlo,janse}, we model swollen polymer chains in solution as self-avoiding walks (SAWs) comprising $n+1$ monomers on a simple cubic lattice. This lattice has a spacing $a=1$ and is effectively infinite in all directions, but it is bounded from below by an impenetrable surface composed of terraces (forming a square lattice of adsorption sites). These terraces have a length $\mathcal{L} \gg n$ and a width $\mathcal{W}$, and are separated by parallel steps. Each step has a height $\mathcal{H}$ and features kinks separated by a distance $\mathcal{D}$ [see Fig.~\ref{fig1}(a)]. The physical scale of interest here is such that the polymer's persistence length and the step height are both comparable to the lattice spacing, meaning that $\ell=\mathcal{H}=a$. One end of the polymer is grafted at a step edge, in the center of the step.

As illustrated in Fig.~\ref{fig1}(b), each monomer residing on the terrace is assigned a statistical weight $\kappa_T = e^{\varepsilon_T/k_B T}$, while those located at the step edge \textit{also} carry a weight $\kappa_S=e^{\varepsilon_S/k_B T}$, where $k_B$ is the Boltzmann constant and the energies $\varepsilon_S$ and $\varepsilon_T$ are both negative. Monomers positioned at kink sites make two distinct contacts with the step, thus contributing a weight of $\kappa_S^2$. Let $C_n(m_S, m_T)$ denote the total number of SAWs of length $n$ having exactly $m_S$ step contacts and $m_T$ terrace contacts. The partition function of the system can then be written as $Z_n(\kappa_S,\kappa_T) = \sum_{m_S,m_T}C_n (m_S,m_T) \kappa_S^{m_S} \kappa_T^{m_T}$, and the expected value of any physical observable $Q$ is given by $\langle Q \rangle(\kappa_S,\kappa_T)=\frac{1}{Z_n}\sum_{\psi_n}\kappa_S^{m_S(\psi_n)}\kappa_T^{m_T(\psi_n)}Q(\psi_n)$, where the sum is evaluated over all possible SAW configurations $\psi_n$. We estimated these thermodynamic averages by performing extensive numerical simulations using the stochastic growth algorithms PERM~\cite{PERM} and flatPERM~\cite{flatPERM}. In both approaches, a chain is grown from an initial monomer (located at the origin of the coordinate system), with each subsequent monomer being added to a randomly chosen, unoccupied neighboring site of the previous one. FlatPERM and PERM both employ pruning and enrichment strategies to reduce attrition and control the weight dispersion during the SAW growth process. The flatPERM algorithm provides direct estimates of $C_n(m_S, m_T)$, which enables us to thoroughly explore the full parameter space for short-length SAWs ($n=128$ here). Conversely, we utilize PERM to perform a more focused analysis at fixed values of $\kappa_S$ and $\kappa_T$, which allows us to study much longer chains ($n=10240$ here). For each parameter set, $\sim 10^9$ walks were generated.

\begin{figure}[t]
    \centering
    \includegraphics[width=8cm]{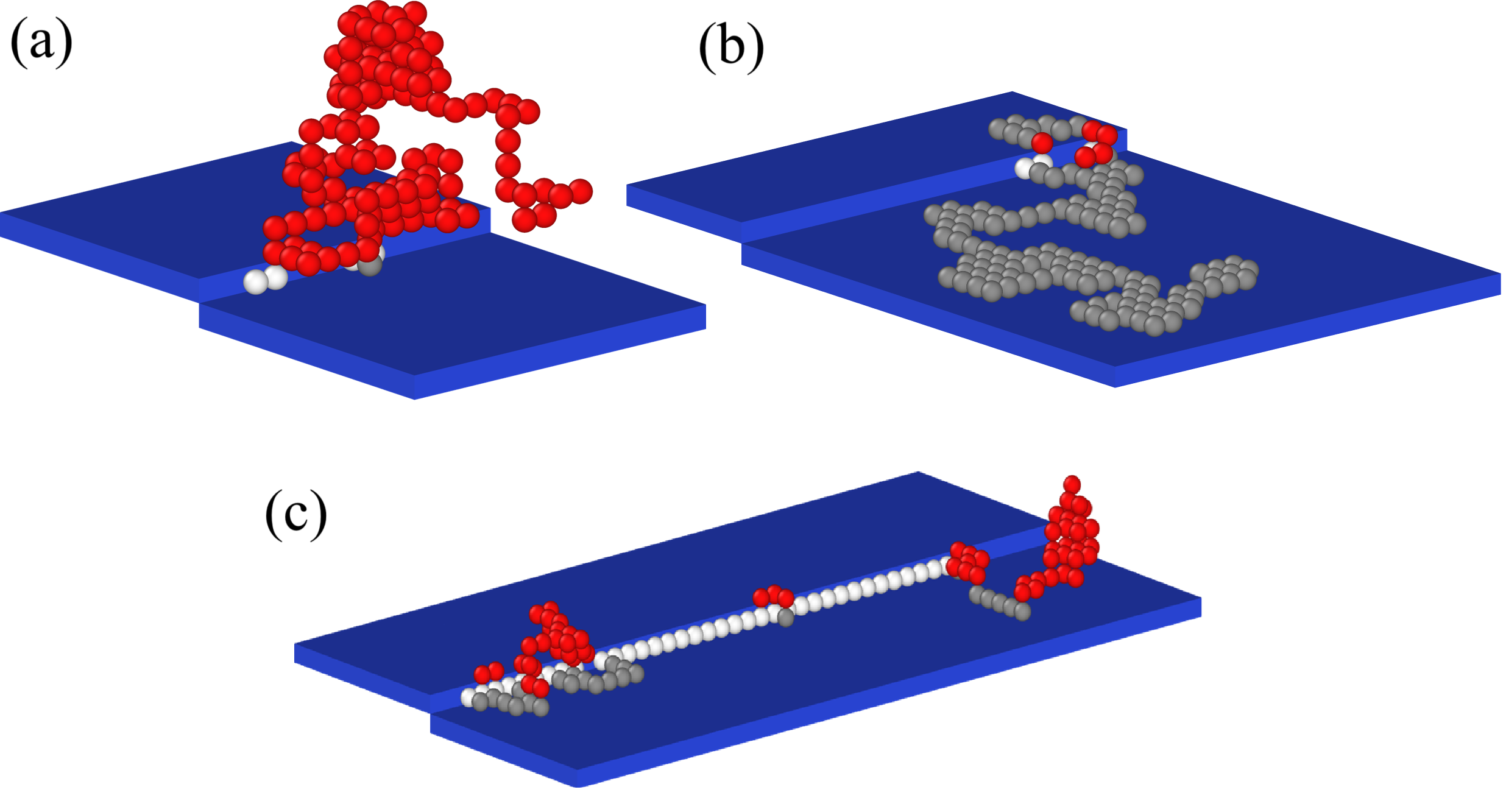}
    \caption{Examples of polymer configurations for $n=128$ in the (a) desorbed ($\kappa_T=\kappa_S=1$), (b) terrace-adsorbed ($\kappa_T=4, \kappa_S=1$), and (c) step-attached ($\kappa_T=2, \kappa_S=5/2$) phases. Red, gray, and white spheres represent desorbed, terrace-adsorbed, and step-attached monomers, respectively.}
    \label{fig2}
\end{figure}


{\bf \textit{Wide terraces with straight steps.}} We begin by discussing the results for straight steps separating broad terraces, ensuring that the polymer interacts with only a single step. This regime corresponds to $\mathcal{D} \gg n$ and $\mathcal{W} \gg R_g^{(2D)}$, where $R_g^{(2D)} \simeq A n^{\nu_{2D}}$ is the average radius of gyration, $\nu_{2D} = 3/4$ is the Flory exponent for 2D SAWs \cite{Flory}, and the amplitude is $A = 0.3289$ for a square lattice \cite{Jensen}. As illustrated in Fig.~\ref{fig2}, three distinct stable phases are identified within the phase diagram:

 \textit{i)} For small values of $\kappa_T$ and $\kappa_S$, the system is characterized by small averages for both $\langle m_T \rangle$ and $\langle m_S \rangle$. This corresponds to the \textit{desorbed} phase, where the polymer chain spans the 3D space [Fig.~\ref{fig2}(a)].

 \textit{ii)} For large $\kappa_T$ and small $\kappa_S$, $\langle m_T \rangle$ is high, but the chain exhibits very few step contacts. Hence, the polymer adopts quasi-2D conformations in this \textit{terrace-adsorbed} (TA) phase [see Fig.~\ref{fig2}(b)].

 \textit{iii)} For large $\kappa_S$, $\langle m_S \rangle$ is large, while the difference $\langle m_T \rangle - \langle m_S \rangle$ remains small, meaning that most adsorbed monomers reside at the step edge. As shown in Fig.~\ref{fig2}(c), the chains are quasi-1D in this \textit{step-attached} (SA) phase.

This three-phase scenario is confirmed both by fluctuation maps for short chains --- which correspond to density plots of the largest eigenvalue of the covariance matrix for $m_S$ and $m_T$, and yield approximations for the phase diagrams of the system; see Fig.~S6 of the Supplemental Material (SM) \cite{SM} --- and by various other observables obtained from PERM simulations for long chains. For instance, a reliable way to determine the adsorption transition lines is via the scaling of the perpendicular and parallel components of the mean squared end-to-end distance, $\langle R^2 \rangle$. In our specific case, however, we must analyze the directional components:
\begin{equation}
    \langle R^2_{\alpha} \rangle(n) \equiv \langle \alpha^2_n\rangle \sim n^{2\nu_{\alpha}},
\end{equation}
where $\alpha_n = x_n, y_n, z_n$ are the coordinates of the free end of the chain, as defined in Fig.~\ref{fig1}(c). In the asymptotic $n \rightarrow \infty$ limit, we expect to find: $\nu_x = \nu_y = \nu_z >0$ in the desorbed phase; $\nu_x = \nu_y > 0$ and $\nu_z=0$ in the TA phase; and $\nu_y > 0$ with $\nu_x=\nu_z=0$ in the SA phase.

\begin{figure}[t]
    \centering
    \includegraphics[width=4cm]{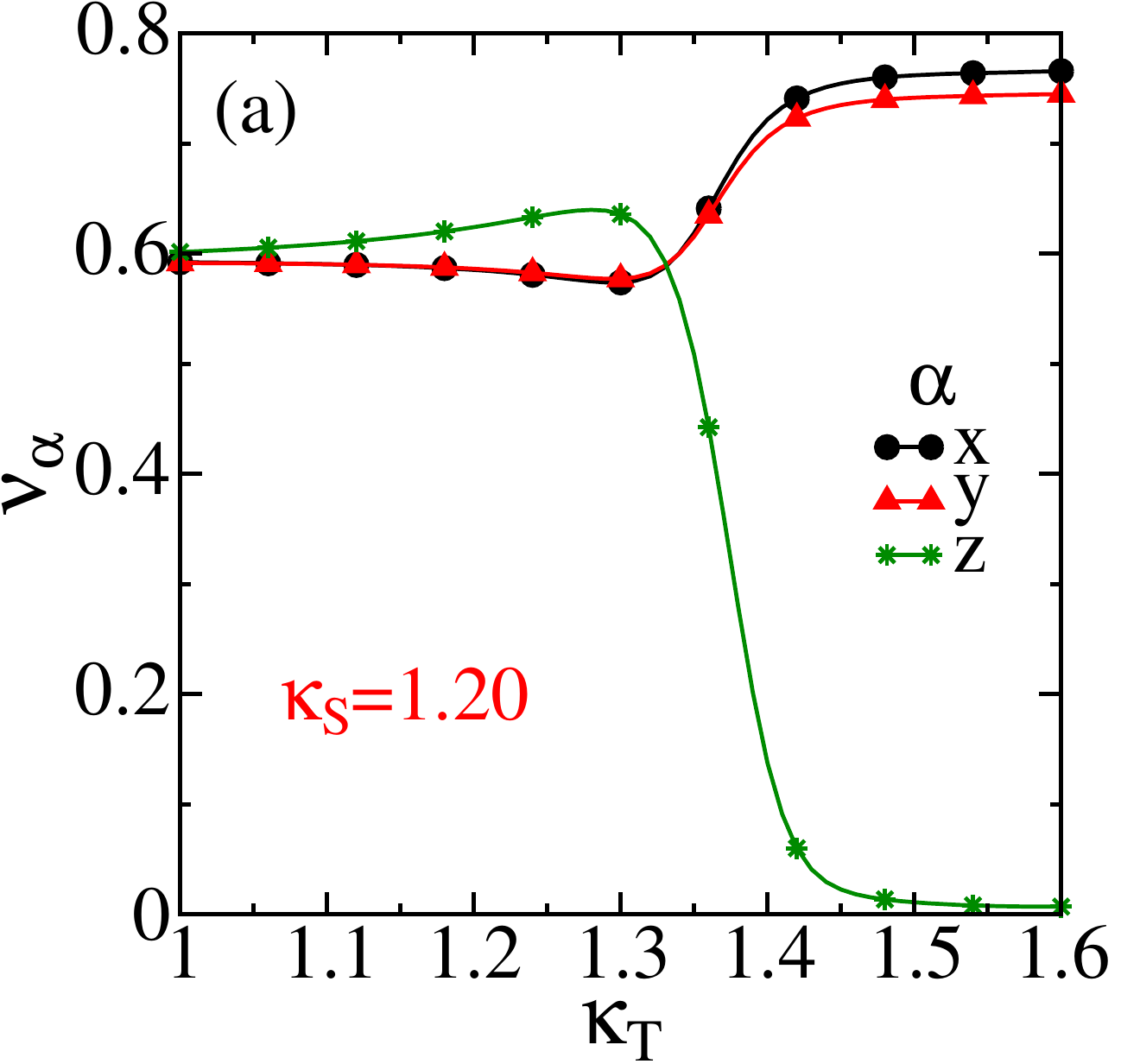}
    \includegraphics[width=4cm]{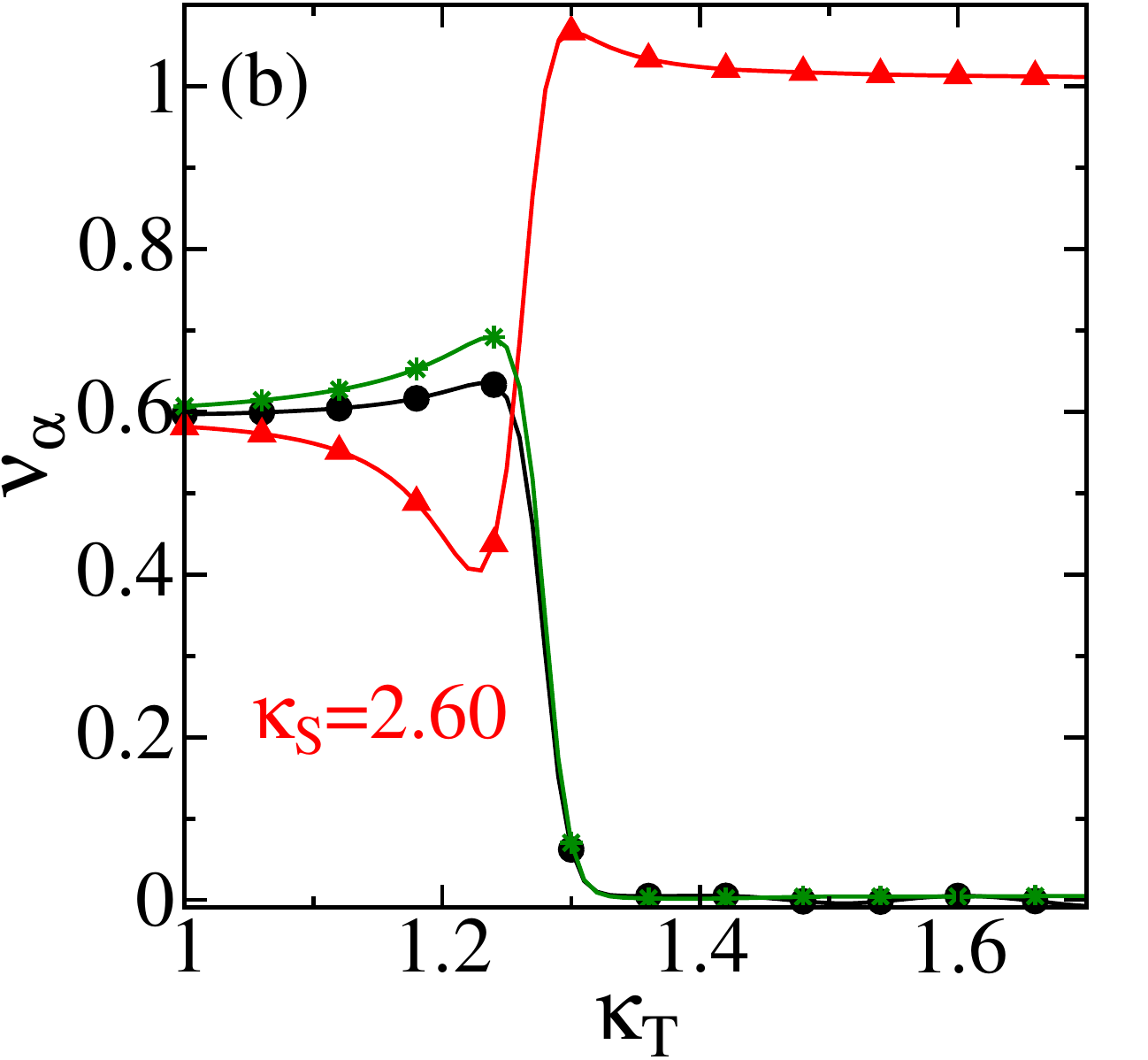}
    \includegraphics[width=4cm]{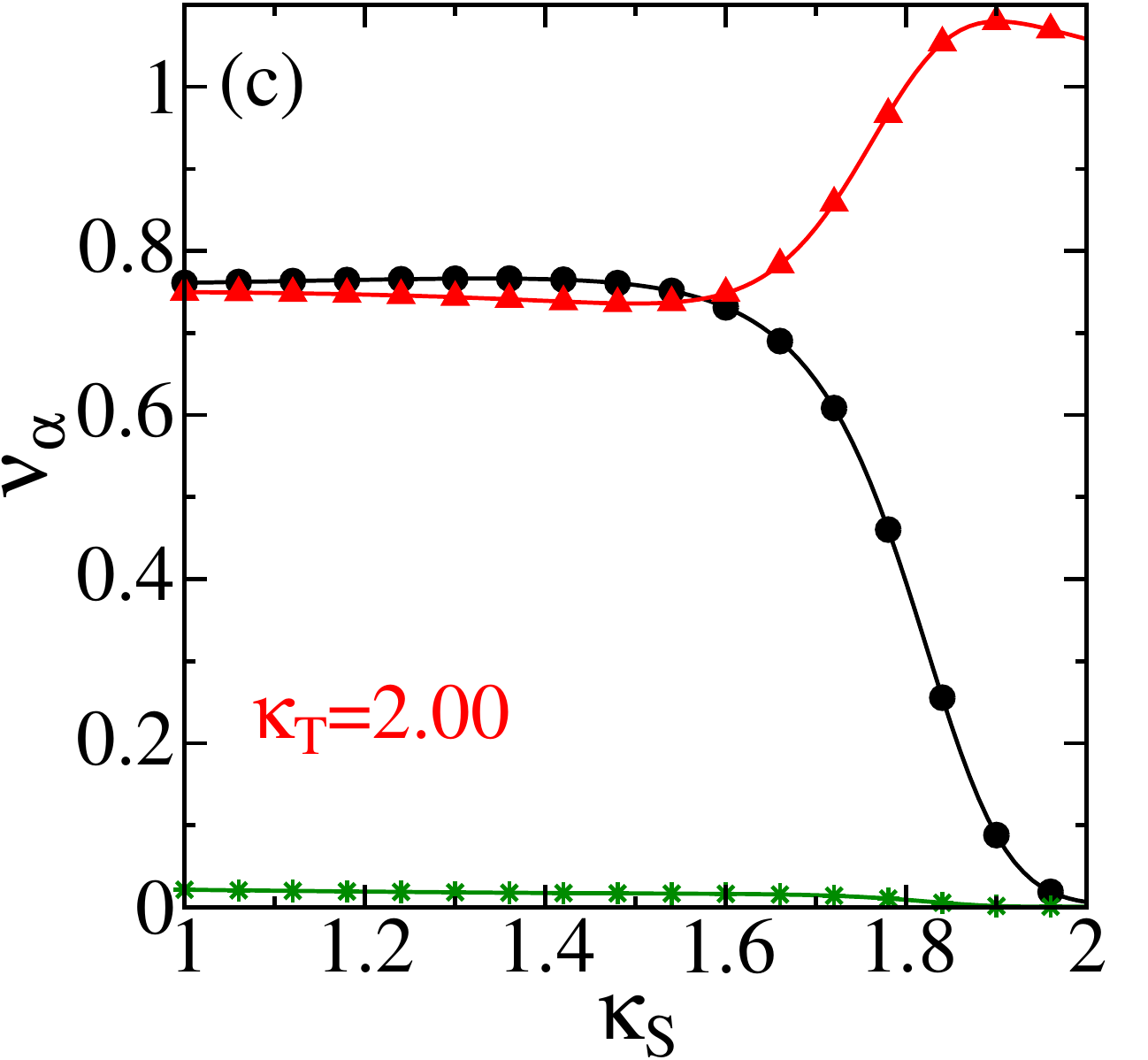}
    \includegraphics[width=4cm]{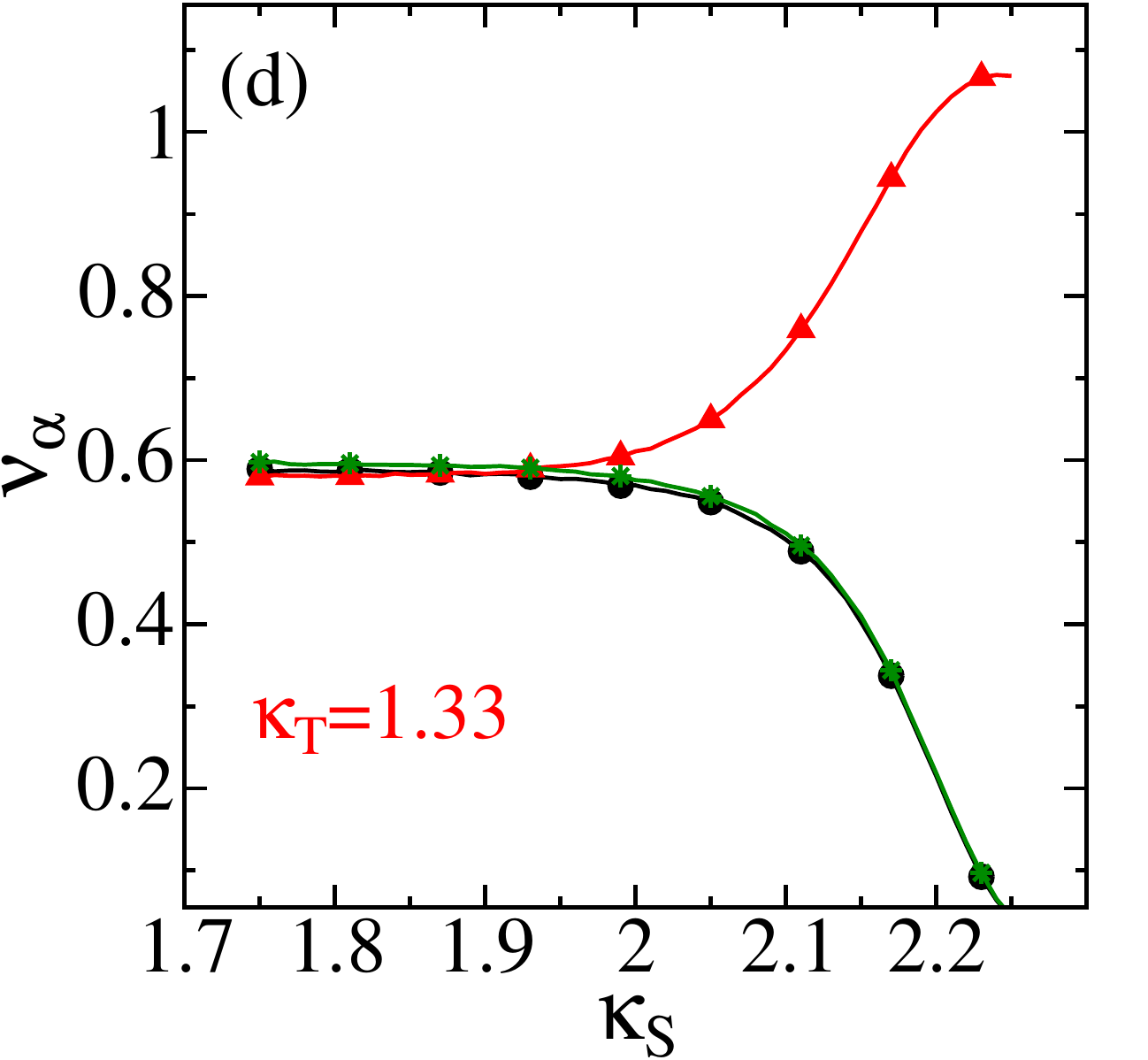}
    \caption{Effective Flory exponents versus $\kappa_T$ ($\kappa_S$) for fixed values of $\kappa_S$ ($\kappa_T$) on a vicinal surface with wide terraces and straight steps. The chain lengths are (a)--(c) $n=1024$ and (d) $n=10240$.}
    \label{fig3}
\end{figure}

\begin{figure}[t]
    \centering
    \includegraphics[width=4cm]{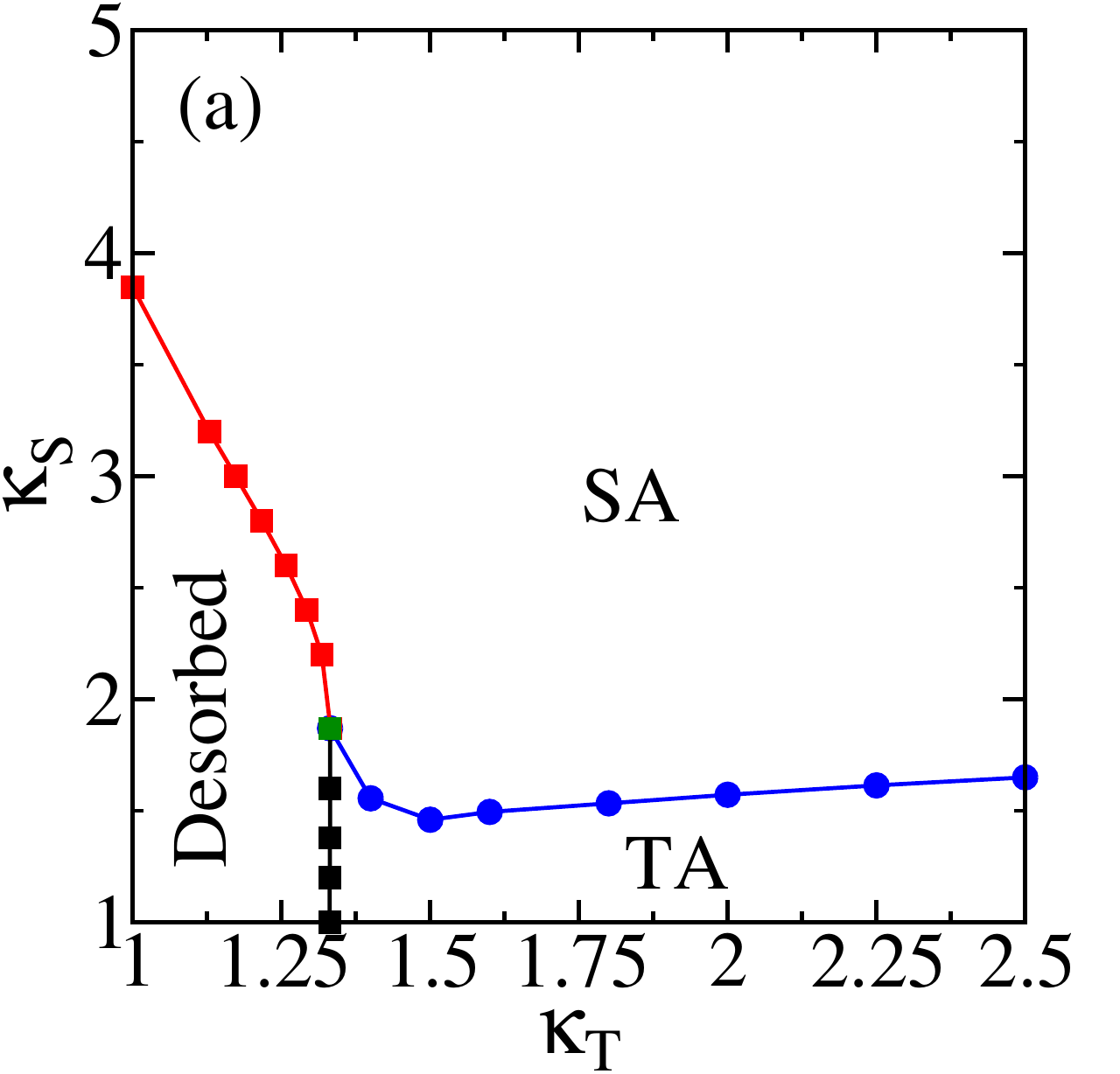}
    \includegraphics[width=4cm]{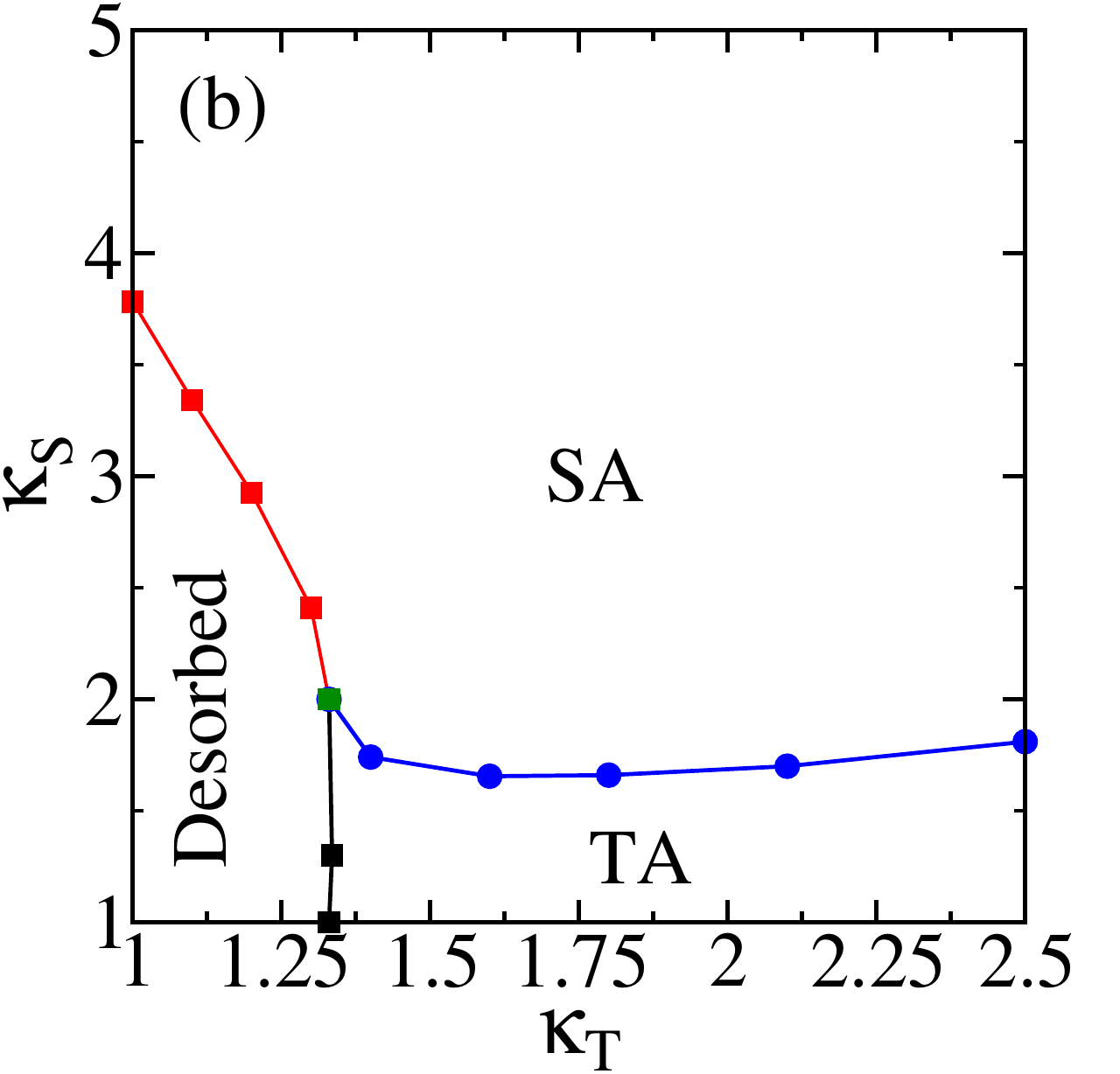}
    \caption{Phase diagrams in the $\kappa_S$--$\kappa_T$ plane for polymer chains tethered on vicinal surfaces with wide terraces separated by (a) straight ($\mathcal{D} \gg n$) and (b) kinked ($\mathcal{D} = 1$) steps.}
    \label{fig4}
\end{figure}

Figure~\ref{fig3}(a) shows that when $\kappa_S$ is small, all exponents $\nu_{\alpha}$ take values close to that for SAWs on 3D lattices ($\nu_{3D} \approx 0.588$ \cite{Clisby}) at small $\kappa_T$, but subsequently shift to $\nu_x \approx \nu_y \approx \nu_{2D} = 3/4$ and $\nu_z \approx 0$ for large $\kappa_T$. This indicates that a phase transition from the desorbed to the TA phase occurs at $\kappa_{T,c} \approx 1.33$, corresponding to the crossing point of the exponent curves. The exact same behavior is observed for $1.0 \leq \kappa_S \lesssim 1.9$ with an identical transition point within our precision, establishing that a desorbed-to-TA transition line exists in the phase diagram at $\kappa_{T,c} \approx 1.33$ [see Fig.~\ref{fig4}(a)]. Notably, this matches the critical point for adsorption on a perfectly flat 2D surface~\cite{Grasberger2005}, which remains unaltered by the presence of the step provided that $\kappa_S \lesssim 1.9$. Further evidence supporting this scenario, derived from other quantities, is presented in the SM, where we also demonstrate that the critical exponents $\phi_{T,c}$ and $1/\delta_{T,c}$ (see Sec.~II and Fig.~S1 of the SM) calculated along this line are entirely consistent with those for flat substrates, i.e., $\phi_{T,c} = 1/\delta_{T,c} \approx 0.48$~\cite{Grasberger2005,Chris1}.

For $\kappa_S \gtrsim 1.9$, we find $\nu_x \approx \nu_y \approx \nu_z \approx \nu_{3D}$ when $\kappa_T$ is small. For large $\kappa_T$, however, Fig.~\ref{fig3}(b) demonstrates that $\nu_y \approx 1$ while $\nu_x \approx \nu_z \approx 0$, as expected for the SA phase. Therefore, increasing $\kappa_T$ at fixed values of $\kappa_S \gtrsim 1.9$ drives the system through a desorbed-to-SA transition. An analogous behavior occurs if one holds $\kappa_T < 1.33$ fixed and increases $\kappa_S$. The resulting critical line is depicted in the phase diagram of Fig.~\ref{fig4}(a), having a maximum value ($\kappa_{S,c} = 3.85$) at $\kappa_T=1$, above which only the SA phase remains stable. This novel transition is confirmed via other thermodynamic quantities in Sec.~III-B of the SM, where the critical exponents $\phi_{S,c} = 1/\delta_{S,c} = 0.86(5)$ are estimated. Although this large value might hint at a first-order transition, no such evidence is observed in the histograms for the number of step contacts $P(m_S)$ [Fig.~S2, SM], strongly suggesting that the desorbed-to-SA transition is continuous.

Fixing $\kappa_T > 1.33$ and increasing $\kappa_S$, as shown in Fig.~\ref{fig3}(c), yields $\nu_x \approx \nu_y \approx \nu_{2D}$ and $\nu_z \approx 0$ for small $\kappa_S$, and $\nu_y \approx 1$ with $\nu_x \approx \nu_z \approx 0$ for large $\kappa_S$. Hence, the system undergoes a TA-to-SA transition. From the intersections of these Flory exponents, we map the corresponding transition line starting at a multicritical point (where it meets the other two critical lines mentioned above) and extending toward $\kappa_T \rightarrow \infty$. In this asymptotic limit, the polymer becomes fully surface-adsorbed, recovering the well-studied case of 2D polymer adsorption onto a 1D boundary line; this transition is continuous, occurring at $\kappa_{S,c} \approx 1.77$ with critical exponents $\phi_{S,c}=1/\delta_{S,c} = 1/2$ \cite{Guim1989}. These same exponents, as well as consistent estimates for the critical points, are observed here for all $\kappa_T \gtrsim 3.0$. For smaller $\kappa_T$, however, the exponents measured conventionally begin to deviate from these expected values, and these deviations grow more pronounced as the multicritical point is approached [Fig.~S3, SM]. This behavior occurs because a considerable fraction of the monomers desorbs when $\kappa_T \lesssim 2.0$, a fraction that systematically increases with decreasing $\kappa_T$. Indeed, by substituting the chain length $n$ with the actual number of adsorbed monomers $n_a \equiv \langle m_T \rangle$ within the scaling relations describing the adsorption, we recover values of $\phi_{S,c}$ and $1/\delta_{S,c}$ that fall significantly closer to $1/2$ [Fig.~S3, SM].

Such pronounced finite-size effects prevent us from accurately locating the multicritical point and its associated critical exponents. Nonetheless, assuming it resides at $\kappa_{T,mc} = 1.33$, as indicated by the results above, the crossing of $\nu_{\alpha}$ versus $\kappa_S$ along this line yields an estimated value of $\kappa_{S,mc} \approx 1.87$ [see Fig.~\ref{fig3}(d)].


{\bf \textit{Wide terraces with kinked steps.}} To investigate the effect of kinks on polymer adsorption, we consider vicinal surfaces with $\mathcal{D} < n$ and $\mathcal{W} \gg R_g^{(2D)}$. We start focusing on the limiting situation $\mathcal{D} = a = 1$. The coordinate system is defined here analogously to Fig.~\ref{fig1}(c), with the $x$- and $y$-directions being perpendicular and parallel to the step, respectively, and $z$ perpendicular to the terrace. The Flory exponents for kinked steps [Fig.~S4, SM] exhibit variations very similar to those displayed in Fig.~\ref{fig3} for the kink-free case. Results for other observables, including the critical exponents, are also analogous to those discussed above for straight steps. Overall, the phase diagrams for polymer adsorption on surfaces with straight and kinked steps with $\mathcal{D}=1$ are almost identical [see Fig.~\ref{fig4}(b)]. A slight upward shift is observed only in the TA-to-SA transition line for the kinked case, as well as in the multicritical point, which is located at $(\kappa_{T,mc},\kappa_{S,mc}) \approx (1.33,2.0)$. The same picture is obtained for $1 < \mathcal{D} < n$, but with intermediate TA-to-SA transition lines and $1.87 \lessapprox \kappa_{S,mc} \lessapprox 2.0$, demonstrating that kinks exert a minor effect on the adsorption process.


{\bf \textit{Short terraces with straight steps.}} Let us thus return to straight steps, considering vicinal surfaces with small terrace widths $\mathcal{W}$ such that the polymer can cross and attach to different steps. Once again, we consider the polymer grafted at a step edge, although similar results are found for short terraces if we move its initial position to the middle of a terrace.

When $\mathcal{W}=a=1$, which corresponds to step bunches forming a $45^\circ$-sloped staircase, every monomer at the surface simultaneously makes one contact with a terrace and one with a step. Consequently, the SA and TA phases become indistinguishable, yielding a single desorbed-to-adsorbed transition, as illustrated by the fluctuation map for $\mathcal{W}=1$ [Fig.~S5, SM]. This transition line occurs at $\kappa_{S,c}\kappa_{T,c} \approx 1.87$, along which the critical exponents are $\phi_{X,c}=1/\delta_{X,c} \approx 0.48$ for $X=T,S$ [Fig.~S5, SM]. Remarkably, even though the polymer temporarily loses contact with the surface when crossing a step, its scaling behavior remains identical to that of perfectly flat surfaces.

Fluctuation maps for widths $\mathcal{W}=2, 4, 8$, and $16$ [Fig.~S6, SM] indicate that the results for $\mathcal{W} = 2$ are highly similar to those for $\mathcal{W} = 1$, though the boundary of the desorbed phase becomes asymmetric with respect to the variables $\kappa_S$ and $\kappa_T$. For $\mathcal{W} = 4$, this boundary is nearly identical to those obtained for larger widths, and a bright stripe separating the regions corresponding to the TA and SA phases emerges in the map. However, the fluctuations remain quite mild along this stripe, indicating a crossover between these two phases, although these fluctuations get stronger as $\mathcal{W}$ increases.

\begin{figure}[t]
    \centering
    \includegraphics[width=4cm]{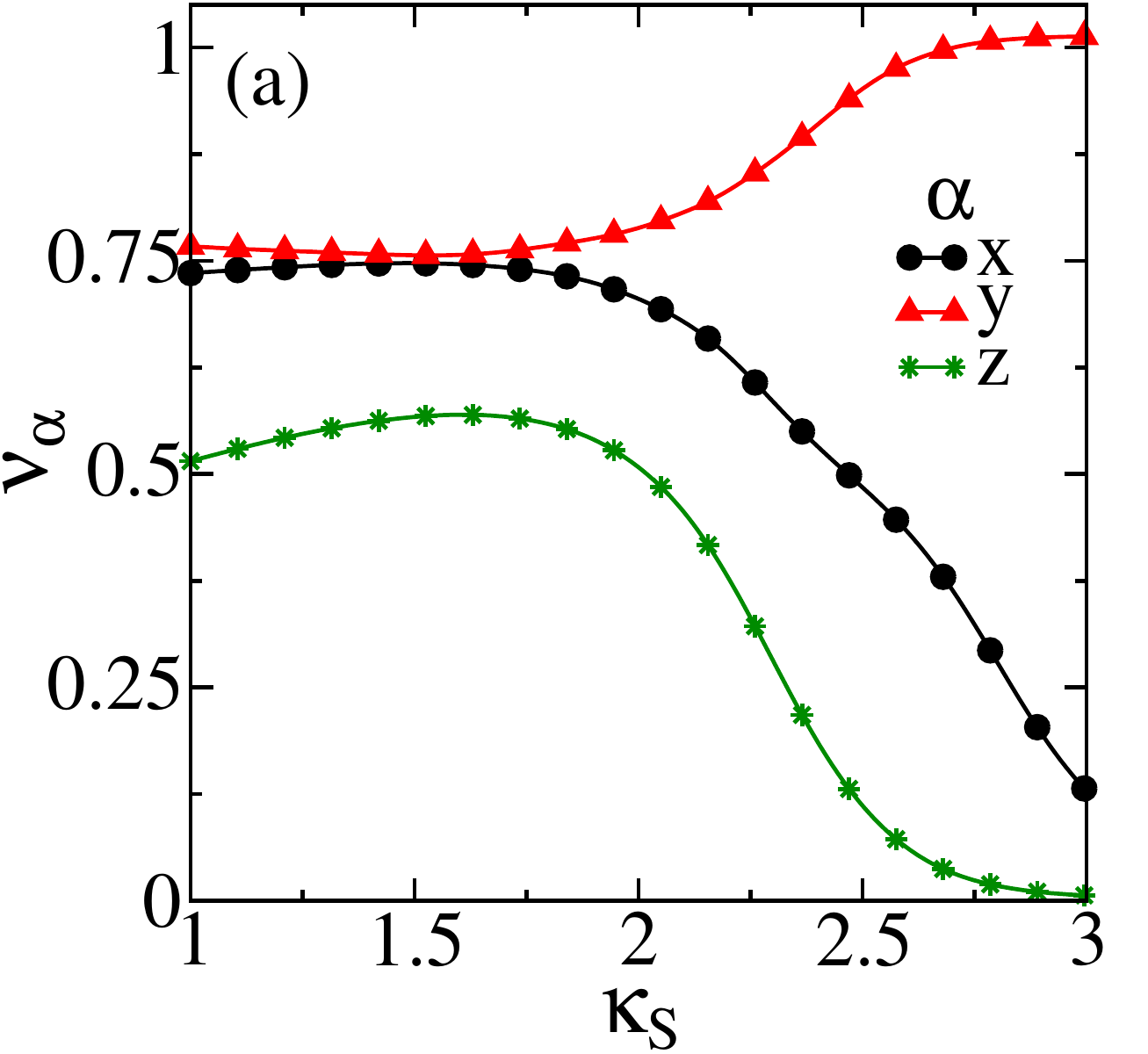}
    \includegraphics[width=4cm]{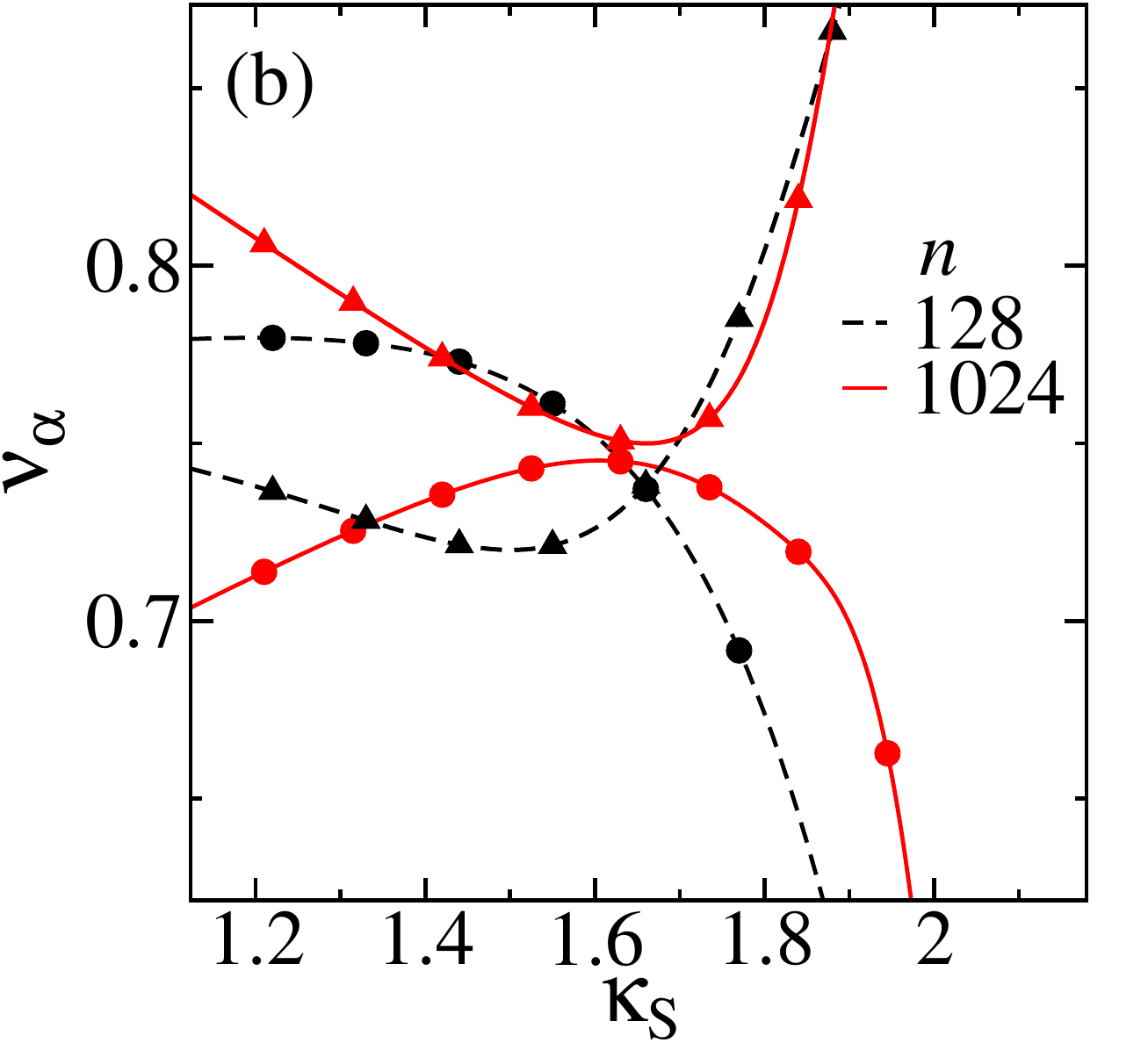}
    \caption{Effective Flory exponents versus $\kappa_S$, for $\kappa_T=2.0$. Data for $\mathcal{W}=19$ and $n=1024$ are shown in (a), while panel (b) presents results for a wider terrace $\mathcal{W}=64$ and different chain lengths. The symbols in (b) follow the legend in (a).}
    \label{fig5}
\end{figure}

It is important to note that keeping $n$ fixed while increasing $\mathcal{W}$ causes the polymer to visit fewer steps, a situation analogous to keeping $\mathcal{W}$ fixed while decreasing $n$. Therefore, to unveil the asymptotic adsorption behavior for short terraces, the finite-size analysis must be performed by keeping the number of visited steps $N_v$ fixed. Equivalently, we discuss this behavior here in terms of the ratio $\mathcal{W}/R_g^{(2D)} \sim 1/N_v$. For instance, Fig.~\ref{fig5}(a) presents the variation of the Flory exponents with $\kappa_S$ for $\kappa_T=2.0$, $n=1024$, and $\mathcal{W}/R_g^{(2D)} \approx 0.32$ (matching the case for $n=128$ and $\mathcal{W}=4$). Similarly to Fig.~\ref{fig3}(c), $\nu_x$ and $\nu_y$ exhibit clear signatures of the TA and SA phases at small and large $\kappa_S$, respectively; but there is no crossing of exponents' curves, consistently with a crossover. (Note that $\nu_z$ is non-zero in the TA phase here because the grafting origin is fixed on a single terrace.) Figure~\ref{fig5}(b) compares the same exponent curves for $\mathcal{W}=64$ at lengths $n=128$ ($\mathcal{W}/R_g^{(2D)} \approx 5$) and $n=1024$ ($\mathcal{W}/R_g^{(2D)} \approx 1$). While no crossing is observed in the latter case, a clear intersection is present for $\mathcal{W}/R_g^{(2D)} \approx 5$.

Results for other observables [shown in Fig.~S7, SM] confirm that only a TA-to-SA crossover exists when $\mathcal{W}/R_g^{(2D)} \lesssim 3$. For example, we find that $\phi_S \approx \phi_T \approx 1$ deep inside both the TA and SA phases for such small ratios, implying that the adsorbed chains always maintain a macroscopic number of step contacts; increasing $\kappa_S$ merely causes them to localize on fewer steps without altering their fundamental scaling behavior. Moreover, similar to the Flory exponents discussed above, we find that curves of $\phi_S$ versus $\kappa_S$ for $\mathcal{W} = 64$ exhibit intersection points characteristic of a critical transition for $n \leq 128$ (corresponding to $\mathcal{W}/R_g^{(2D)} \gtrsim 5$). However, the curves for $n=300$ and $n=1024$ (for which $\mathcal{W}/R_g^{(2D)} \approx 2.7$ and $\approx 1$, respectively) do not cross (either each other or the other curves), confirming that the criticality changes to a crossover when $\mathcal{W}/R_g^{(2D)} \lesssim 3$ [Fig.~S7(b), SM]. On the other hand, by increasing $n$ while keeping this ratio fixed at $\mathcal{W}/R_g^{(2D)} \approx 3.5$, for example, the signature of criticality appears for all simulated lengths [Fig.~S7(c), SM]. This strongly suggests that the spatial constraint separating the crossover and critical behaviors is at $\mathcal{W}/R_g^{(2D)} \approx 3$. Additional evidence comes from the exponents $\phi_S$ measured deep inside the TA phase for different ratios $\mathcal{W}/R_g^{(2D)}$ [Fig.~S7(d), SM]. Their effective values stop depending on this ratio and approach zero when $\mathcal{W}/R_g^{(2D)} \gtrsim 3$, demonstrating that this is the condition for the number of monomer-step contacts becoming negligible in the TA phase, allowing for the onset of a critical TA-to-SA transition.

It is important to note that the physical spatial extent of the polymer in the $x$-direction is given by the transverse mean span (caliper size) of the chain, $S_{x}(n) \simeq B n^{\nu_{2D}}$ \cite{janse}. We estimate here the ratio $S_x/R_g^{(2D)} = B/A \approx 2.4$ for 2D SAWs on the square lattice [Fig.~S8, SM], which lies intermediate between the exact value for ideal Gaussian chains, $S_x/R_g^{(2D)} = \sqrt{24/\pi} \approx 2.76$ \cite{Weiss}, and that for self-avoiding rings, $S_x/R_g^{(2D)} \approx 2.12$ \cite{Privman1985}. Consequently, the condition $\mathcal{W} \gtrsim 3 R_g^{(2D)}$ translates directly to $\mathcal{W} \gtrsim 1.2 S_x$, demonstrating that the chain must be strictly confined within a single terrace for the TA-to-SA criticality to manifest.


{\bf \textit{Conclusions.}} Our study reveals three possible thermodynamic scenarios for polymer adsorption on vicinal surfaces, depending on the terrace width: (\textit{i}) a single desorbed-to-adsorbed transition for small $\mathcal{W} \sim a$, analogous to the adsorption transition on flat 2D substrates; (\textit{ii}) the same single transition for intermediate $\mathcal{W}$, but with the adsorbed phase presenting a crossover from the TA to the SA phase for large $\kappa_T$ and increasing $\kappa_S$; and (\textit{iii}) the phase diagrams displayed in Fig.~\ref{fig4} for large $\mathcal{W}$, where the three phases are separated by continuous transition lines. The presence of kinks in the steps does not alter this behavior for the parallel steps analyzed here.

Substantially, regardless of whether it experiences a crossover or a true phase transition, the polymer in scenarios (\textit{ii}) and (\textit{iii}) will ultimately end up in the SA phase at low $T$, provided it interacts attractively with the step. If $\kappa_S = \kappa_T$, starting from the desorbed phase and decreasing $T$ causes the chain to first adsorb onto the terrace and then attach to the step. For other interactions, a direct transition from the desorbed to the SA phase is possible. The desorbed-to-SA transition, alongside the multicritical point discovered in scenario (\textit{iii}), are novel features that invite future theoretical works to explain their underlying critical properties (particularly the large critical exponents observed here). Moreover, this highlights the importance of further investigating polymer adsorption on other experimentally relevant vicinal surfaces --- such as those with non-parallel steps, rough steps, AB terraces, or alternating terrace-step bunches --- as well as on complex crystalline surfaces featuring islands, mounds, or screw dislocations. The interplay between polymer size and terrace width revealed here indicates that polymerization at vicinal surfaces is also an interesting topic for future studies.

From a practical perspective, precise control of polymer conformation is crucial for tuning its electronic, mechanical, and biological properties for specific applications. Therefore, the existence of a stable SA phase, as unveiled here, indicates that such control can be achieved via attachment-detachment on engineered stepped surfaces (e.g., using lithography techniques). Furthermore, vicinal surfaces have long been utilized as templates for the self-assembled growth of nanowires formed by atomic chains decorating the steps \cite{nanowires}. Our results demonstrate that robust polymer nanowires can be produced at vicinal surfaces by setting the thermodynamic parameters deep inside the SA phase. Importantly, this approach can overcome the persistent difficulty in obtaining long wires caused by atomic defects (inherent to self-assembly), which frequently fragment the chains --- since the wires will be formed via adsorption of long, continuous polymers.

\acknowledgements

The authors acknowledge financial support from CNPq and FAPEMIG (Brazilian agencies). This research utilized Queen Mary's Apocrita HPC facility, supported by QMUL Research-IT~\cite{Apocrita}, on which all simulations have been performed.

\end{document}


\title{Supplementary Material for\\
``Polymer adsorption on vicinal surfaces''}

\author{Nathann T. Rodrigues}
\affiliation{International Center of Physics, Institute of Physics, University of Brasilia, 70910-900, Brasilia, Federal District, Brazil} \affiliation{Departamento de F\'isica, Universidade Federal de Vi\c cosa, 36570-900, Viçosa, Minas Gerais, Brazil}
\author{Tiago J. Oliveira}
\affiliation{Departamento de F\'isica, Universidade Federal de Vi\c cosa, 36570-900, Viçosa, Minas Gerais, Brazil}

\date{September 2, 2026}

\maketitle

\makeatletter
\def\l@f@section{}%
\makeatother
\tableofcontents

\section{Introduction}

This Supplementary Material (SM) provides additional details regarding the simulations of polymer chains adsorbing onto vicinal surfaces. We begin by defining the key quantities analyzed in this study. Subsequently, we discuss further data for polymer adsorption on surfaces with straight steps and wide terraces, followed by a complementary analysis of how kink sites and terrace width affect the polymer phase behavior.


\section{Quantities of interest}

To construct the phase diagrams for polymer adsorption, we first examine the fluctuation maps obtained from flatPERM simulations for $n=128$. These maps correspond to density plots of the largest eigenvalue of the covariance matrix:
\begin{equation}
   \begin{bmatrix}
\langle m_T^2 \rangle - \langle m_T \rangle^2 & \langle m_T m_S \rangle - \langle m_T \rangle \langle m_S \rangle \\
\langle m_S m_T \rangle - \langle m_S \rangle \langle m_T \rangle & \langle m_S^2 \rangle - \langle m_S \rangle^2
\end{bmatrix},
\end{equation}
where $m_T$ and $m_S$ are, respectively, the numbers of monomer-terrace and monomer-step contacts. Regions of high fluctuation (indicated by brighter colors in the maps below) typically mark the phase boundaries.

The average fractions of monomer contacts with terraces, $U_{T,n}$, and with steps, $U_{S,n}$, are important quantities for characterizing adsorption transitions. They are defined as:
\begin{equation}
U_{X,n}(\kappa_S,\kappa_T)=\frac{\langle m_X\rangle}{n},
\label{EqUn}
\end{equation}
where $X = T,S$. In the commonly studied scenario of three-dimensional (3D) polymers adsorbing onto a perfectly flat 2D substrate, one has $U_{T,n}=0$ and $U_{T,n}=1$ in the ground states of the desorbed and adsorbed phases, respectively. A similar behavior is observed for $U_{S,n}$ when 2D polymers adsorb onto a 1D boundary. Therefore, in such simple geometries, these fractions serve as the order parameters for the adsorption transitions.

In the vicinity of the adsorption point (e.g., $\kappa_{Y,c}$), they are expected to scale as:
\begin{equation}
    U_{X,n} \sim n^{\phi_X-1}\mathcal{F}_X(\tau n^{1/\delta_X}),
    \label{EqScalingUn}
\end{equation}
where $\tau=\kappa_Y-\kappa_{Y,c}$ and $X,Y = T$ or $S$. The scaling function $\mathcal{F}_X(z)$ assumes a constant value at $z=0$, meaning that the curves of the effective exponents $\phi_{Y}$ versus $\kappa_X$ become independent of $n$ (i.e., they cross) exactly at the critical point when $n \rightarrow \infty$ \cite{Thomas2018}. Note that, depending on the dimensionality of the transition (3D to 2D, 2D to 1D, etc.), only one of the fractions ($U_{T,n}$ or $U_{S,n}$) will exhibit these scaling properties.

Other key observables are the relative fluctuations in the number of terrace contacts, $\Gamma_{T,n}$, and step contacts, $\Gamma_{S,n}$, which are defined as:
\begin{equation}
    \Gamma_{X,n}(\kappa_S,\kappa_T) = \kappa_Y \frac{\partial \log U_{X,n}}{\partial \kappa_Y} = \frac{\langle m_X m_Y \rangle-\langle m_X\rangle \langle m_Y\rangle}{\langle m_X \rangle},
    \label{EqGamma}
\end{equation}
where $X,Y=T$ or $S$. At least one of these fluctuations will display a maximum near the critical point, the peak value of which is expected to scale as:
\begin{equation}
    \Gamma_{X,n}^{(max)} \sim n^{1/\delta_{X}}.
    \label{EqGammaMax}
\end{equation}
Moreover, the location of this maximum provides an estimate for the pseudocritical parameters $\kappa_S$ or $\kappa_T$.


\section{Results for vicinal surfaces with wide terraces and straight steps}

As discussed in the main text, the phase diagram for polymers adsorbing onto surfaces where the steps are straight (i.e., free of kinks) and the terraces are sufficiently wide (ensuring the polymer interacts with only a single step) exhibits three distinct phases. These are: the desorbed phase (at small weights $\kappa_S$ and $\kappa_T$); the terrace-adsorbed phase (at small $\kappa_S$ and large $\kappa_T$); and the step-attached phase (at large $\kappa_S$ and $\kappa_T$).  In the following subsections, we present a detailed analysis of the transition lines separating these phases and their corresponding critical properties.

\subsection{Desorbed to Terrace-Adsorbed Transition}

At low $\kappa_S$, increasing $\kappa_T$ drives the polymer from the desorbed (3D coil) phase to the terrace-adsorbed phase. For perfectly flat substrates, this corresponds to the well-documented 3D to 2D continuous adsorption transition, which occurs at $\kappa_T \approx 1.33$~\cite{Grasberger2005}, with critical exponents $\phi_{c}=1/\delta_{c} \approx 0.48$~\cite{Grasberger2005,Thomas2018}.

\begin{figure}[t!]
    \centering
    \includegraphics[width=4cm]{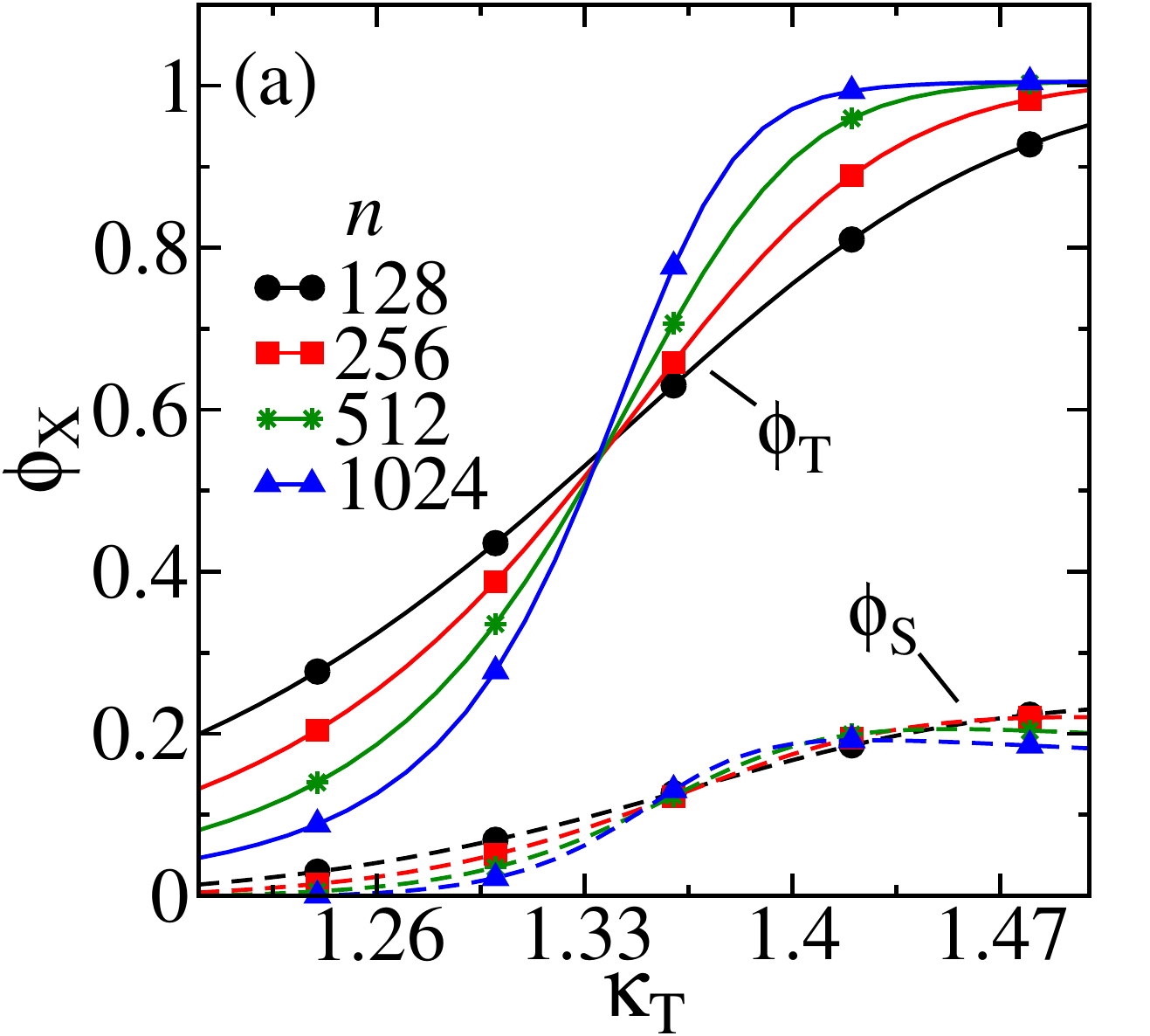}
    \includegraphics[width=4cm]{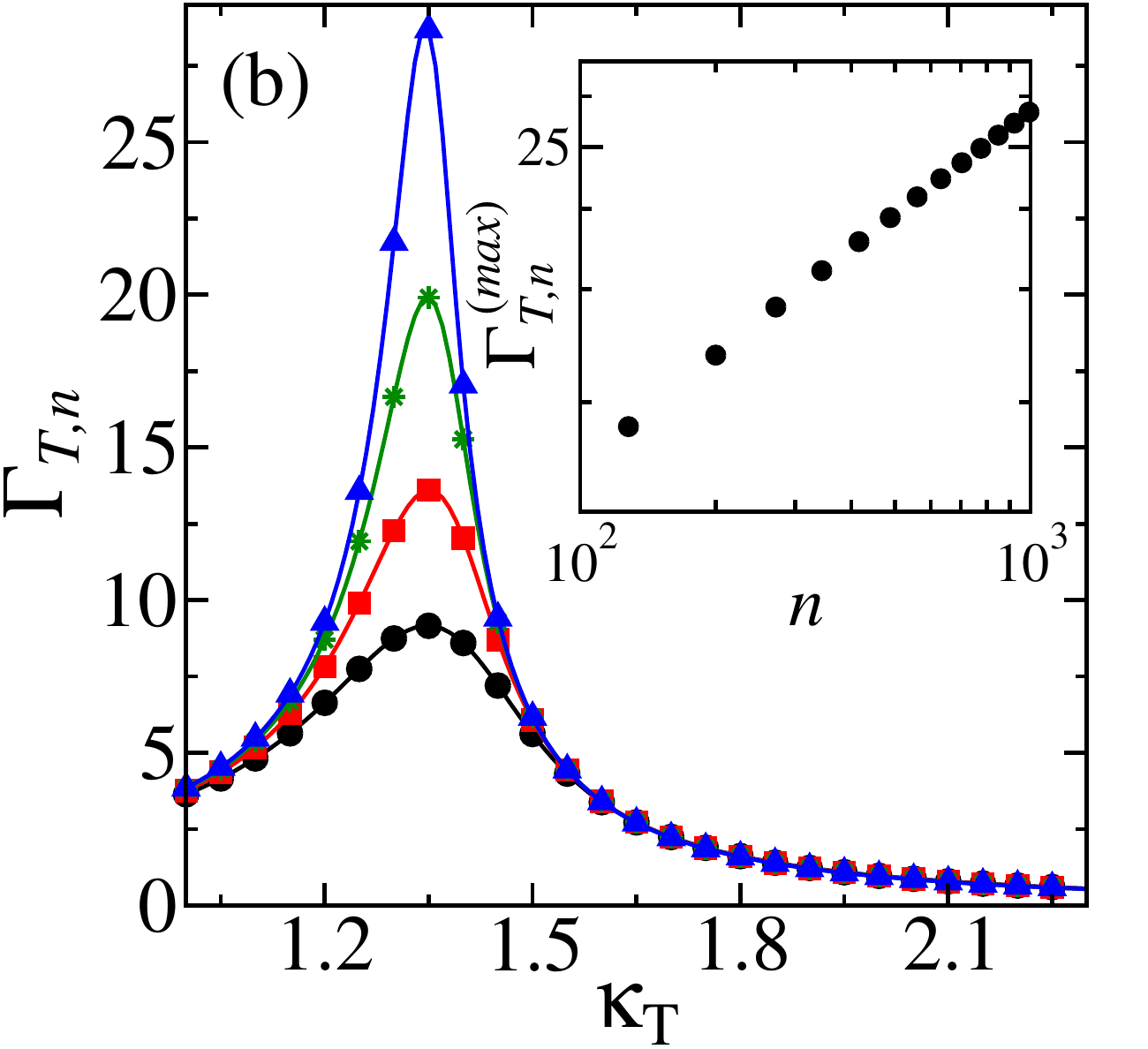}
    \caption{Results for vicinal surfaces with broad terraces and straight steps. (a) Effective exponents $\phi_S$ (dashed) and $\phi_T$ (solid lines), and (b) fluctuations $\Gamma_{T,n}$ as functions of $\kappa_T$, for $\kappa_S = 1.20$. Data for different chain lengths $n$ are shown in both panels, as indicated by the legend in (a). The inset in (b) presents the maxima of $\Gamma_{T,n}$ against $n$.}
    \label{figSM1}
\end{figure}

Figure~\ref{figSM1}(a) displays the effective exponents $\phi_T$ and $\phi_S$ as functions of $\kappa_T$, for $\kappa_S = 1.2$. These effective exponents were calculated as the local slopes of the $U_X$ versus $\kappa_T$ curves, where $X=T,S$. The exponent $\phi_T$ approaches zero for small $\kappa_T$, as expected for the desorbed phase, and increases towards $1$ (indicating polymer adsorption) for large $\kappa_T$. Conversely, $\phi_S$ consistently retains low values, demonstrating that terrace contacts overwhelmingly dominate over step contacts, which remain minimal across both phases. The crossing of the $\phi_T$ curves for different chain lengths $n$ confirms the presence of a continuous adsorption transition at $\kappa_T \approx 1.33$. Furthermore, the value of $\phi_T$ at the crossing point aligns well with $\phi_{T,c} = 0.48$. An identical behavior is observed in the $\phi_T \times \kappa_T$ curves for other values of $\kappa_S \lesssim 1.9$, giving rise to a desorbed to terrace-adsorbed transition line situated at $\kappa_{T,c} \approx 1.33$, a finding also corroborated by the Flory exponents discussed in the main text. Along this line, we estimate critical exponents that fluctuate within the interval $0.48 \leq \phi_{T,c} < 0.53$ for SAWs up to 10240 steps, exhibiting no systematic trend to increase or decrease with varying $\kappa_S$.

The relative fluctuations $\Gamma_{T,n}$ versus $\kappa_T$, for $\kappa_S = 1.2$, are plotted in Fig.~\ref{figSM1}(b); analogous results are found for other values of $\kappa_S \lesssim 1.9$. The progressive build-up of these fluctuations with increasing $n$ provides additional evidence for a continuous transition. Consistent with the other observables, the locations of the $\kappa_T$ maxima systematically converge to $\kappa_T \approx 1.33$ as $n \rightarrow \infty$. Moreover, $\Gamma_{T,n}^{(max)}$ scales as $n^{1/\delta_{T}}$ [see the inset of Fig.~\ref{figSM1}(b)], yielding crossover exponents in the range $0.49 < 1/\delta_{T,c} < 0.51$. This strongly suggests that the relation $\phi_{T,c} = 1/\delta_{T,c}$ holds along the entirety of the desorbed to terrace-adsorbed transition line. Crucially, these exponents remain very close to the established flat-surface values of $\phi_{T,c} = 1/\delta_{T,c} \approx 0.48$~\cite{Grasberger2005,Thomas2018}.

\subsection{Desorbed to Step-Attached Transition}

\begin{figure}[t!]
    \centering
    \includegraphics[width=4cm]{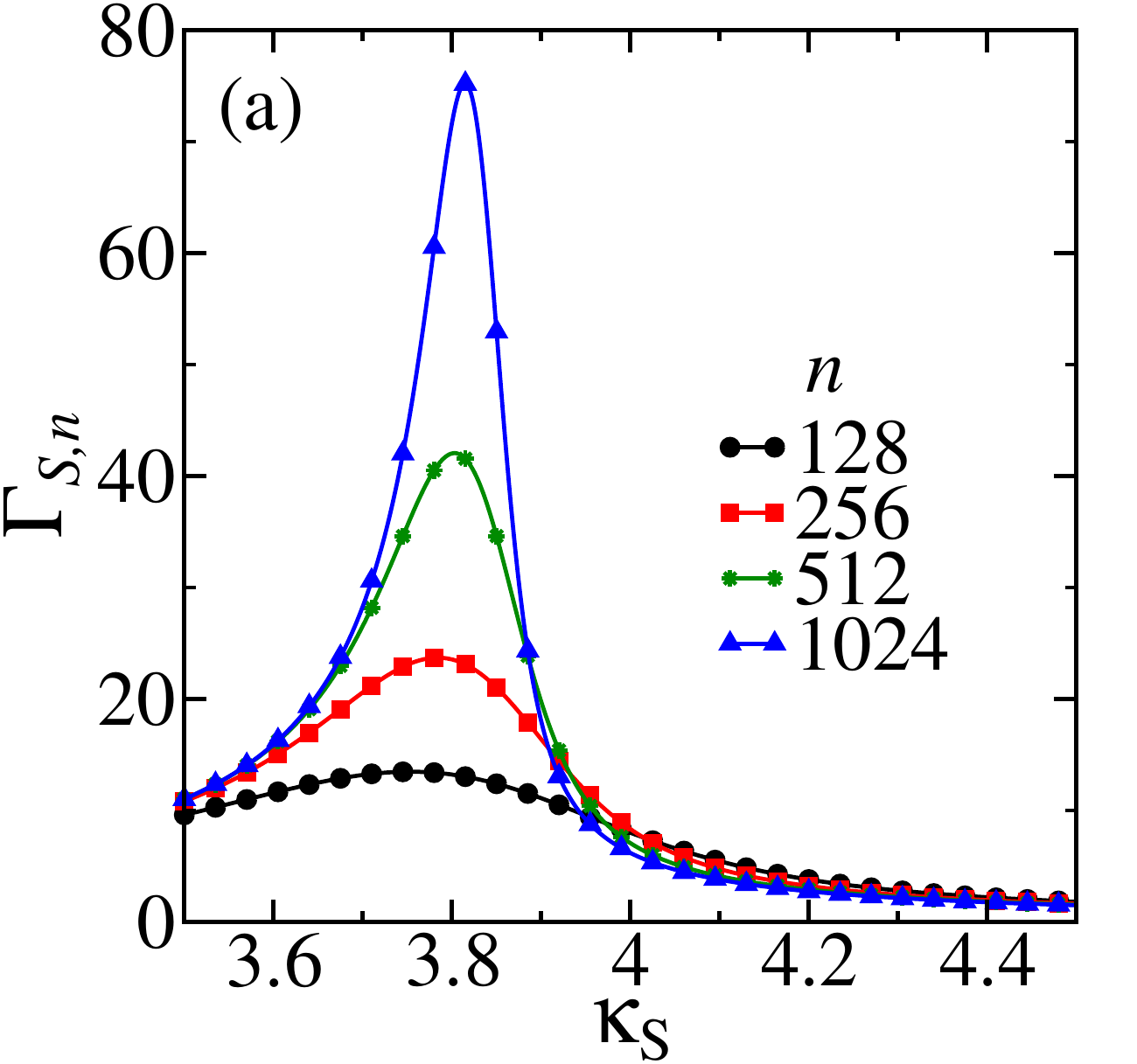}
    \includegraphics[width=4cm]{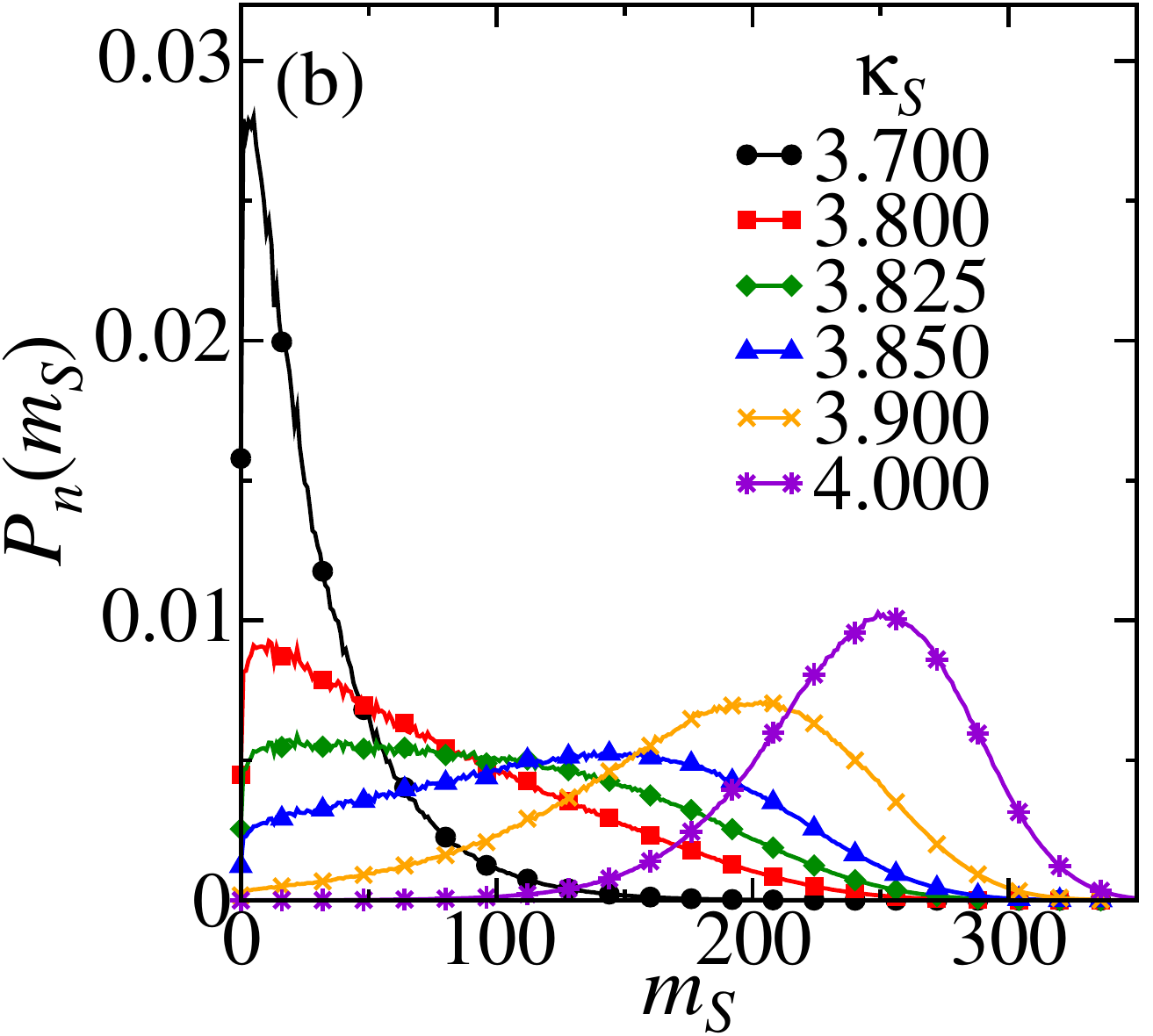}
    \includegraphics[width=4cm]{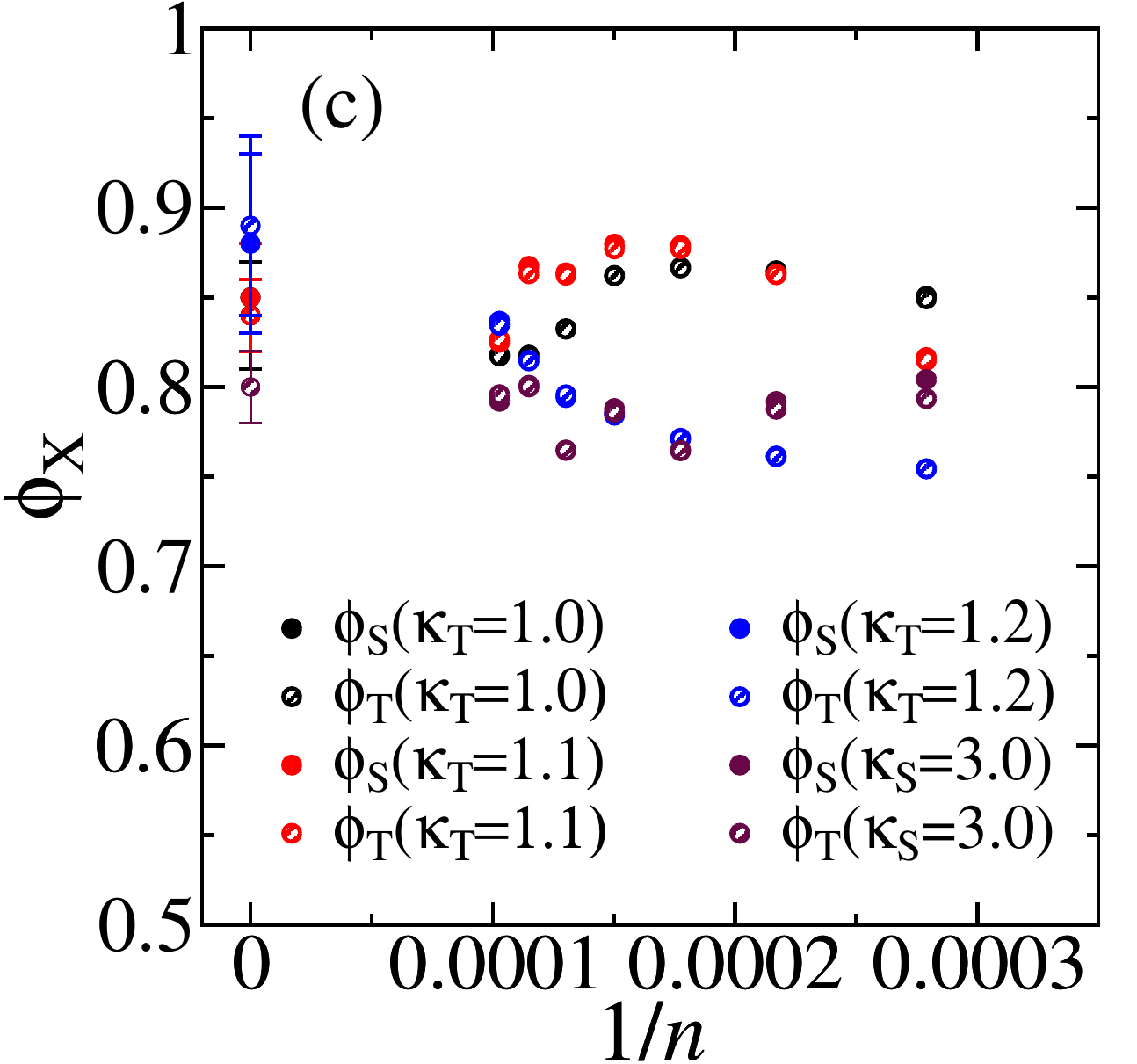}
    \includegraphics[width=4cm]{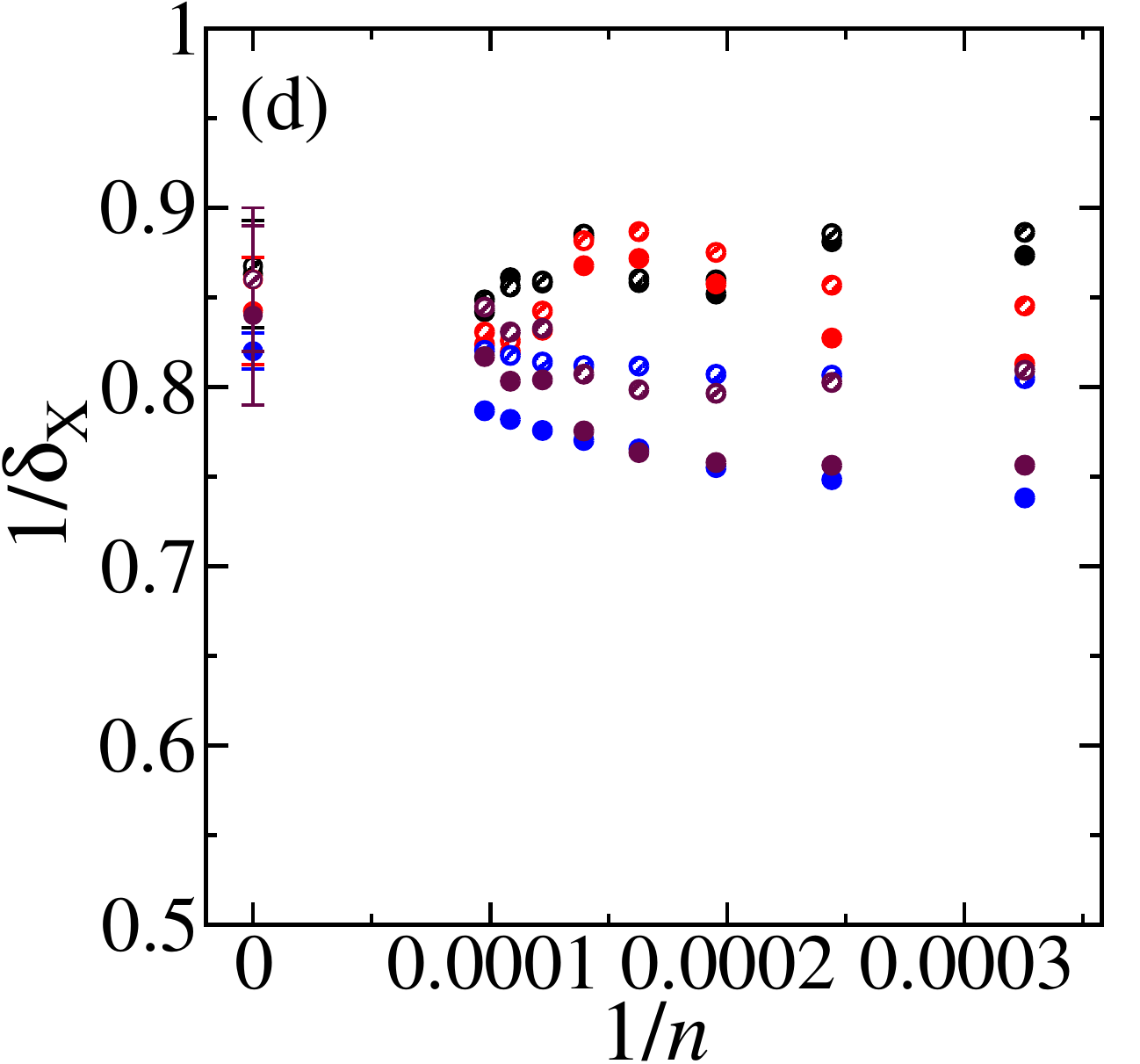}
    \caption{Results for vicinal surfaces with broad terraces and straight steps. (a) Fluctuations $\Gamma_{S,n}$ versus $\kappa_S$, for the indicated lengths $n$. (b) Distribution $P_n(m_S)$ of monomer-step contacts at different values of $\kappa_S$ near the critical point, for $n=512$. The results in panels (a) and (b) are for $\kappa_T = 1.0$, and similar behaviors are found for other $\kappa_T < 1.33$. Panels (c) and (d) show the extrapolations of the pseudocritical exponents $\phi_{T}$, $\phi_{S}$, $1/\delta_{T}$, and $1/\delta_{S}$ obtained along the desorbed to step-attached transition line for fixed parameters indicated in the legend of (c).}
    \label{figSM2}
\end{figure}

For $\kappa_T < 1.33$, increasing $\kappa_S$ drives the system through a transition from the desorbed phase to the step-attached phase. An equivalent transition is observed when $\kappa_S \gtrsim 1.9$ is held constant while increasing $\kappa_T$.

Figure~\ref{figSM2}(a) provides a representative plot of the fluctuations $\Gamma_{S,n}$ versus $\kappa_S$, for $\kappa_T=1.00$. Notably, these fluctuations intensify with $n$ at a much faster rate than those associated with the desorbed to terrace-adsorbed transition shown in Fig.~\ref{figSM1}(b). While such a rapid increase is typically a hallmark of first-order transitions, the corresponding probability distributions $P_n(m_S)$ for the number of monomer-step contacts lack any trace of bimodality near the transition point [see Fig.~\ref{figSM2}(b)]. The monomer-terrace contact distributions $P_n(m_T)$ [not shown] similarly exhibit no bimodal behavior. Thus, the desorbed to step-attached transition appears to be genuinely continuous.

The rapid growth in the fluctuations points to unexpectedly large critical exponents. Because every step contact inherently acts as a terrace contact, the effective exponents $\phi_S$ and $\phi_T$ are almost identical. Indeed, the intersection points of these exponent curves [analogous to those in Fig.~\ref{figSM1}(a), though not shown here] provide nearly equivalent estimates for their pseudocritical values. These estimates, obtained for various $\kappa_T \in [1.0, 1.2]$ as well as for fixed $\kappa_S=3.0$, are summarized in Fig.~\ref{figSM2}(c). Extrapolating these values to the thermodynamic limit ($n \rightarrow \infty$) yields $\phi_{S,c} \approx \phi_{T,c} = 0.87(6)$. In cases where the data did not demonstrate monotonic convergence, only the largest chain lengths were utilized for the extrapolation. The critical points derived from these curve intersections are also in strong agreement with the transition points estimated independently via the Flory exponents and the $\Gamma_{S,n}$ maxima. The resulting desorbed to step-attached transition boundary is mapped out in Fig.~2(a) of the main text.

As anticipated, the crossover exponents are similarly large. Figure~\ref{figSM2}(d) demonstrates that the finite-size values $1/\delta_S(n)$ and $1/\delta_T(n)$ occupy the same range as $\phi_S(n)$ and $\phi_T(n)$. Upon extrapolating this data, we estimate the thermodynamic crossover exponents to be $1/\delta_{S,c} \approx 1/\delta_{T,c} = 0.85(4)$. This evidence strongly implies that the scaling relation $\phi_{X,c} = 1/\delta_{X,c} \approx 0.86$ is maintained for both $X=T$ and $S$ along this newly identified transition line.

\subsection{Terrace-Adsorbed to Step-Attached Transition}

For $\kappa_T > 1.33$, the system undergoes a transition from a terrace-adsorbed state to a step-attached state as $\kappa_S$ increases. In the extreme limit of $\kappa_T \rightarrow \infty$, the polymer lies entirely flat on the surface, reducing the scenario to the `classic' 2D to 1D continuous adsorption transition, known to occur at $\kappa_S \approx 1.77$ with exact critical exponents $\phi_{S,c} = 1/\delta_{S,c} = 1/2$ \cite{Guim1989}.

\begin{figure}[t!]
    \centering
    \includegraphics[width=4cm]{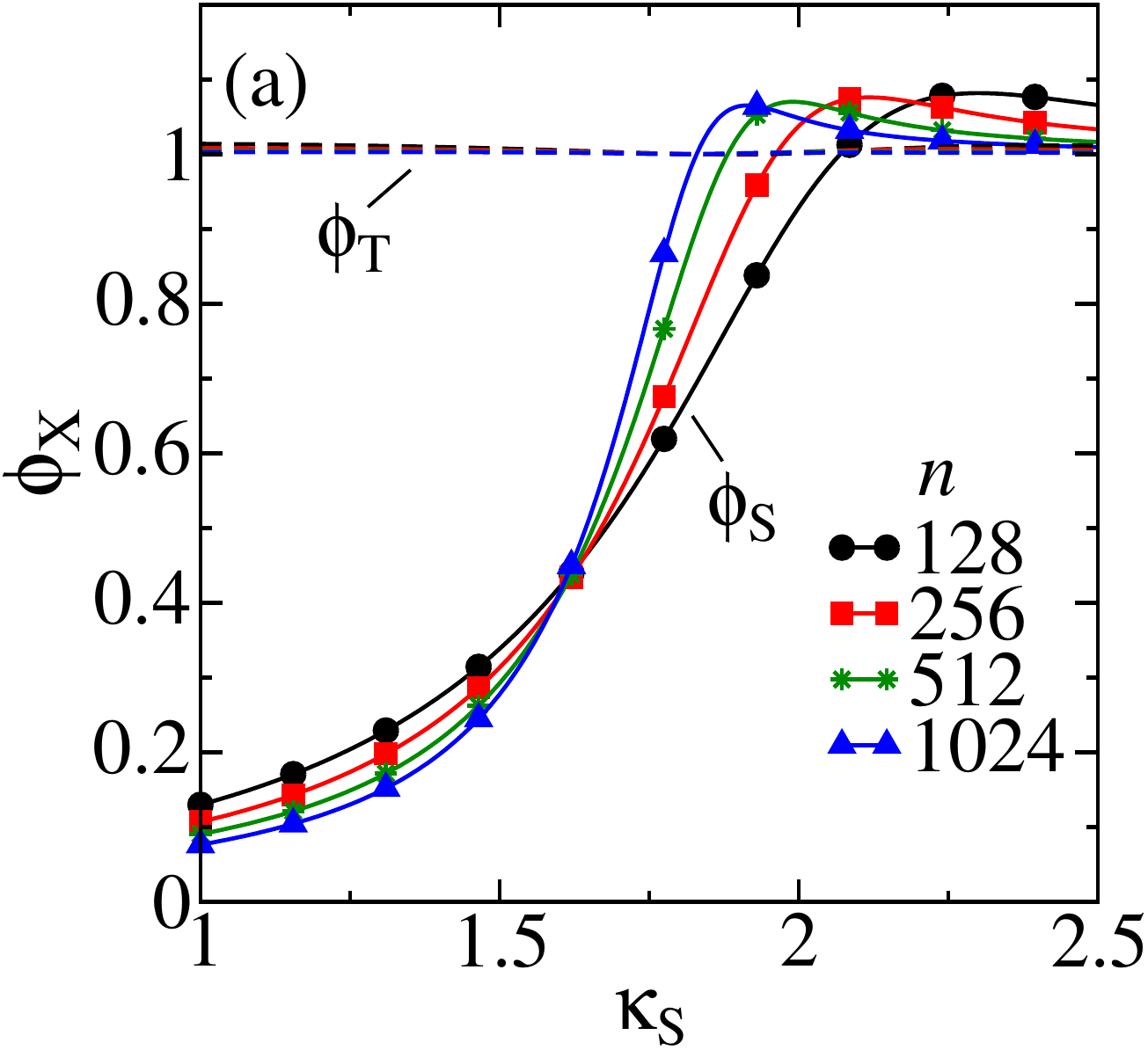}
    \includegraphics[width=4cm]{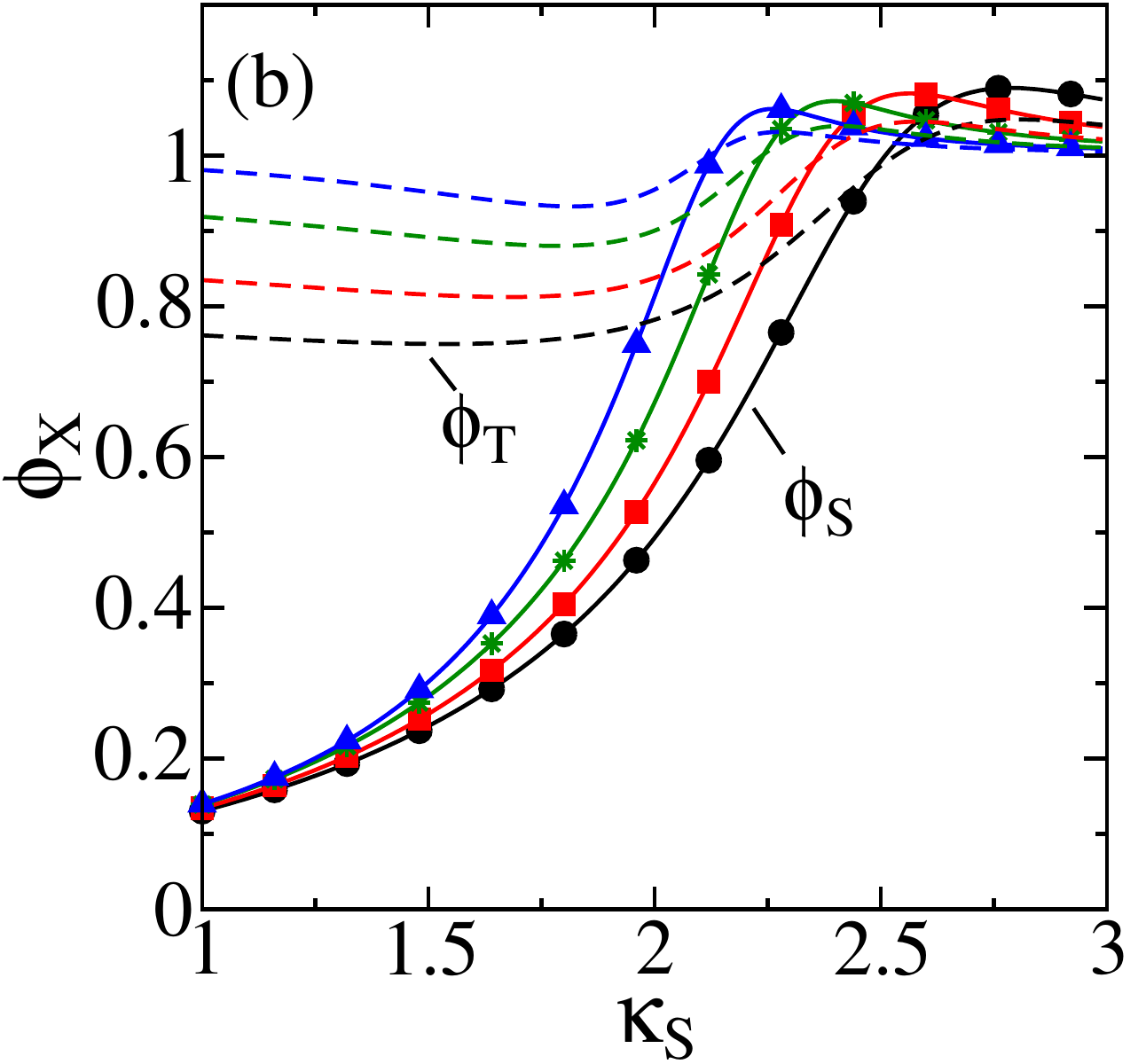}
    \includegraphics[width=4cm]{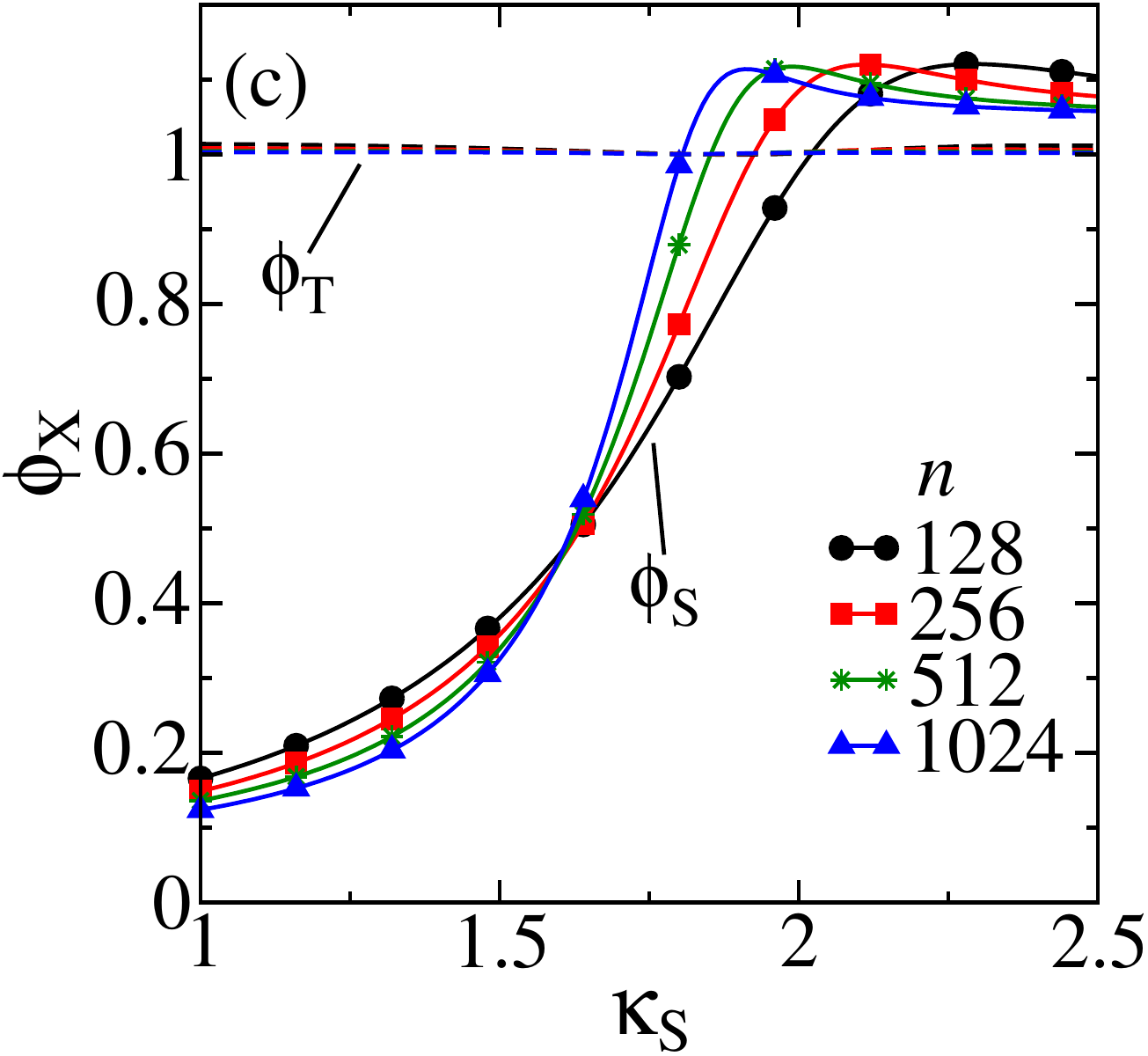}
    \includegraphics[width=4cm]{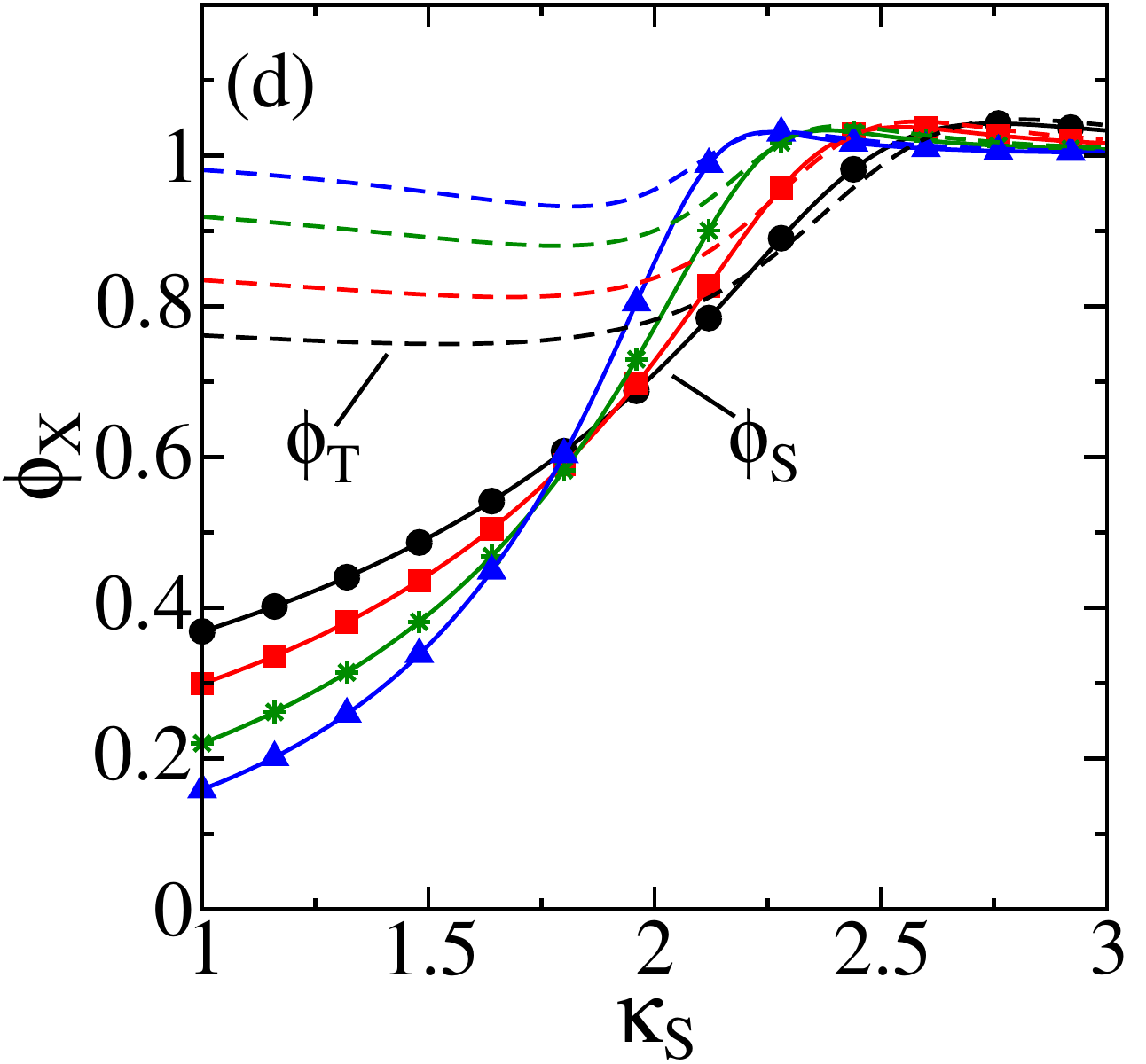}
    \caption{Results for vicinal surfaces with broad terraces and straight steps. Effective exponents $\phi_T$ (dashed lines) and $\phi_S$ (solid lines) plotted as a function of $\kappa_S$, evaluated at $\kappa_T = 2.0$ [(a) and (c)] and $\kappa_T = 1.4$ [(b) and (d)], for the chain lengths listed in the legend of (a). The exponents in the top panels were calculated using the standard scaling relation in Eq.~\ref{EqScalingUn}, whereas the bottom panels utilized the modified scaling in Eq.~\ref{EqPhiNew}.}
    \label{figSM3}
\end{figure}

We observe a highly analogous scenario here for $\kappa_T \gtrsim 3.0$; however, the behavior becomes more intricate at smaller values of $\kappa_T$. For instance, at $\kappa_T = 2.0$, Fig.~\ref{figSM3}(a) shows that $\phi_T \approx 1$ across all values of $\kappa_S$, suggesting the vast majority of monomers are already adsorbed to the terrace. Yet, at the crossing point, $\phi_S \approx 0.4$, which is significantly smaller than the theoretical $1/2$. As $\kappa_T$ is lowered further, the values of $\phi_S$ at the crossings continue to drop, and these intersection points disappear entirely as $\kappa_T$ approaches the multicritical point, as evidenced in Fig.~\ref{figSM3}(b) for $\kappa_T = 1.4$. Although a cursory glance might suggest the absence of a terrace-adsorbed to step-attached transition in this regime, we argue that this disappearance is purely an artifact of severe finite-size effects.

Specifically, the average fraction of \textit{desorbed} monomers (defined as $1-\langle m_T \rangle/n$) is approximately $23\%$ at the transition point for $\kappa_T = 2.0$, but surges to roughly $80\%$ for $\kappa_T = 1.4$. Crucially, the scaling relation in Eq.~\ref{EqScalingUn} is physically expected to hold for $U_{S,n}$ only when the polymer is rigidly constrained to a 2D space, rendering $U_{S,n}$ the true fraction of step contacts per adsorbed monomer. In our actual 3D system, the properly normalized equivalent fraction is:
\begin{equation}
 U^*_{S,n}(\kappa_S,\kappa_T) = \frac{\langle m_S \rangle}{n_a},
 \label{EqUnalt}
\end{equation}
where $n_a \equiv \langle m_T \rangle$ represents the average number of surface-adsorbed monomers. By analogy with Eq.~\ref{EqScalingUn}, it naturally follows that:
\begin{equation}
 U^*_{S,n} \sim n_a^{\phi_S-1} f_a(\tau n_a^{1/\delta_S}).
 \label{EqPhiNew}
\end{equation}
When the effective exponents $\phi_S(n)$ are recalculated using this metric and plotted against $\kappa_S$, the expected curve crossings unequivocally reappear for all values of $\kappa_T$. Furthermore, these intersections occur at values of $\kappa_S$ that match the critical points extracted independently from the Flory exponents. Even more compellingly, the values of $\phi_S$ at these new crossings correctly converge toward $1/2$, as demonstrated in Figs.~\ref{figSM3}(c) and \ref{figSM3}(d).

The finite-size distortions caused by the desorbed monomers also heavily impact the normalized fluctuations. For example, if we attempt to estimate the crossover exponent $1/\delta_S$ from the maxima of the uncorrected $\Gamma_S$ curves [defined in Eq.~\ref{EqGamma}] versus $\kappa_S$, we extract unphysically small values near the multicritical point (e.g., $1/\delta_S \approx 0.1$ for $\kappa_T=1.4$). Additionally, the critical points derived from these naive maxima fail to align with those established by other observables.

\begin{figure}[t]
    \centering
    \includegraphics[width=4cm]{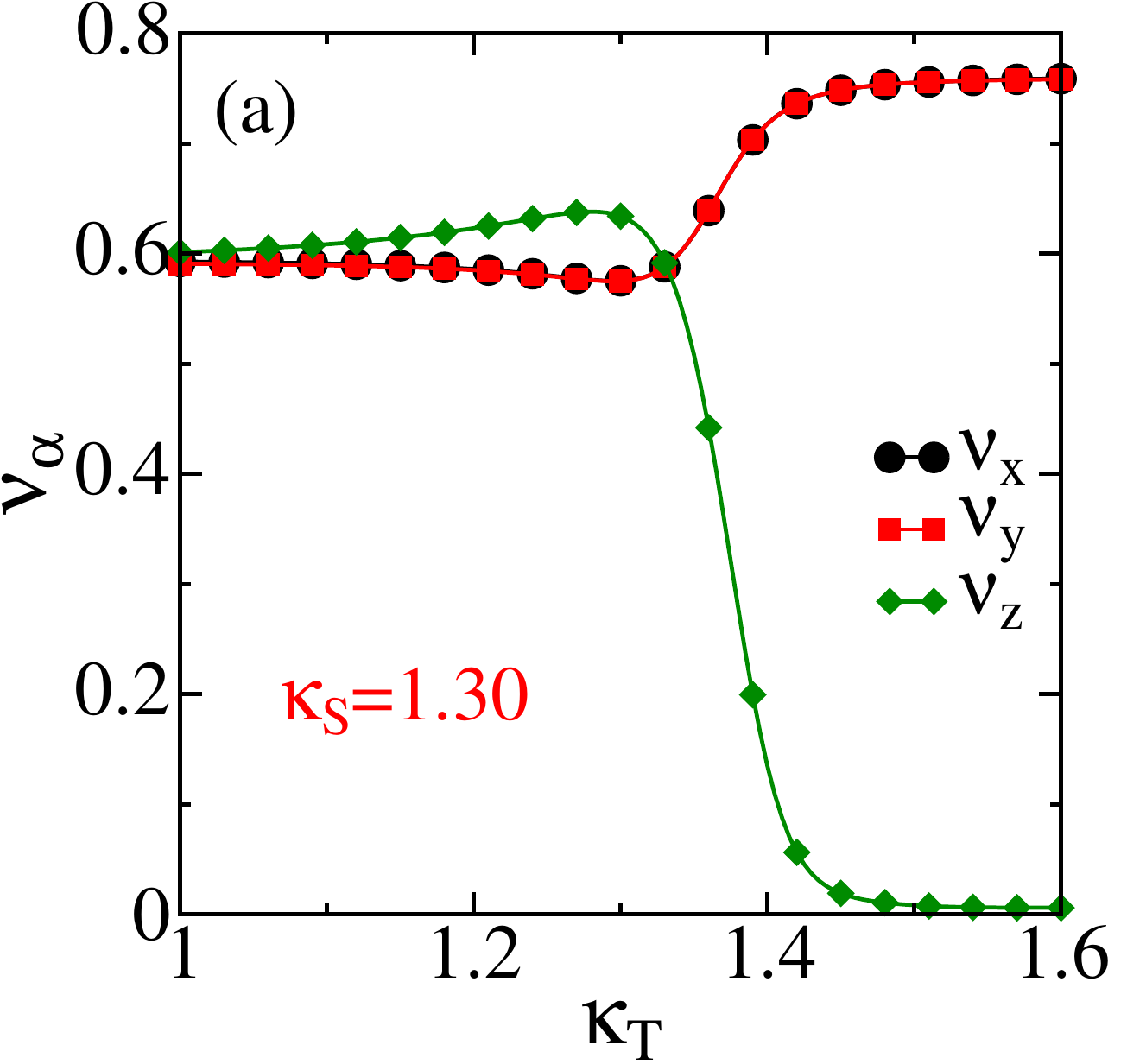}
    \includegraphics[width=4cm]{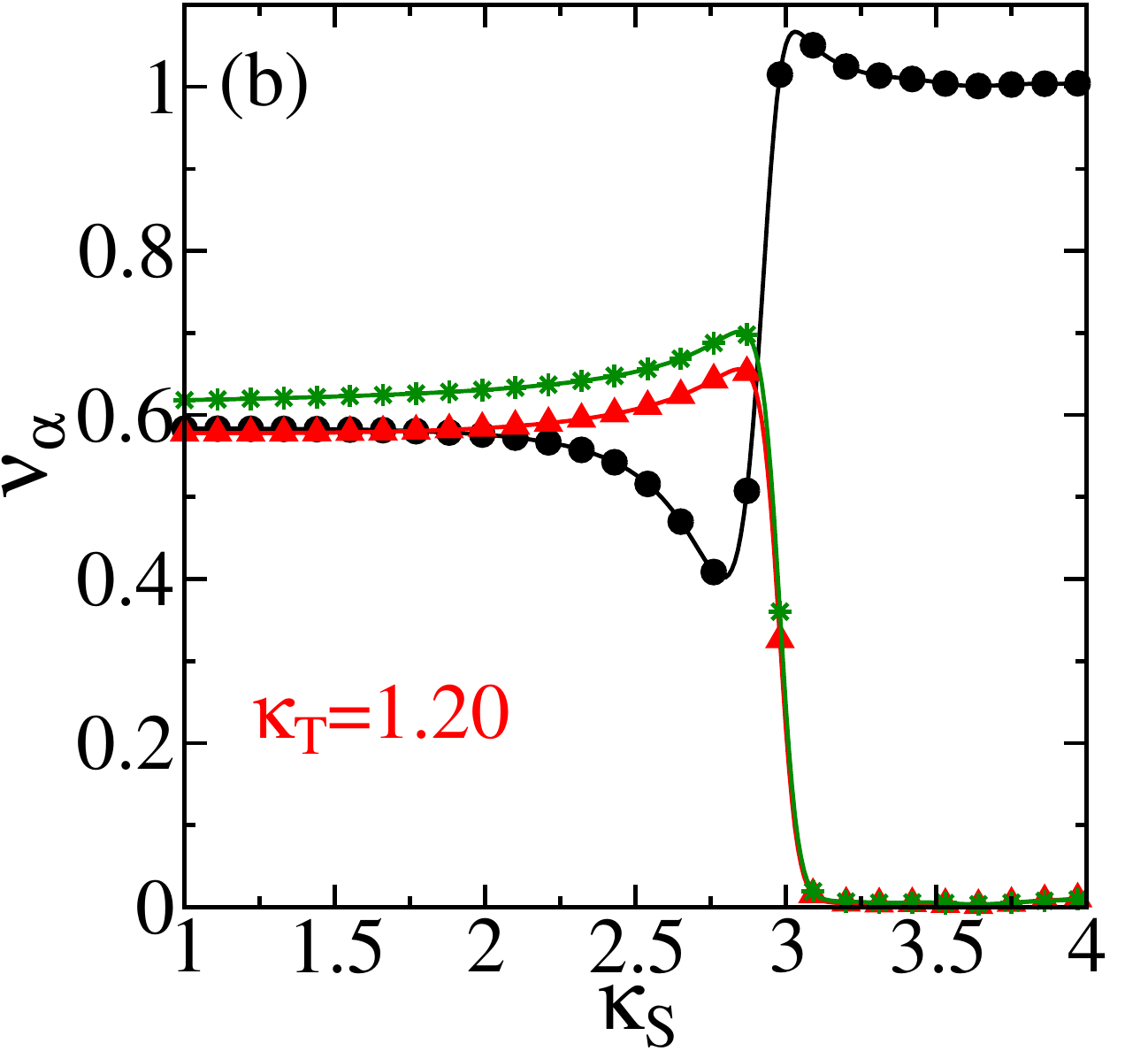}
    \includegraphics[width=4cm]{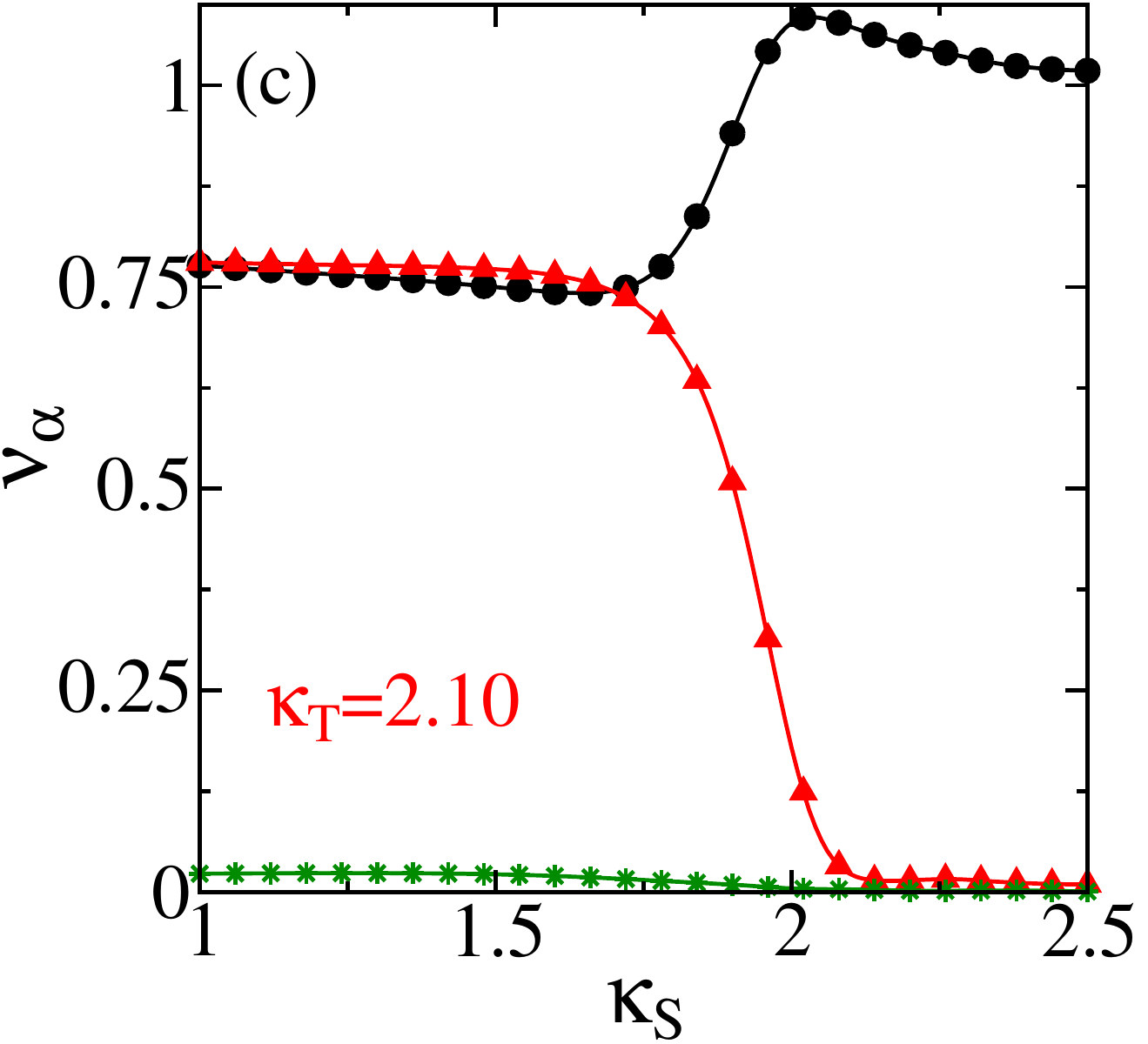}
    \includegraphics[width=4cm]{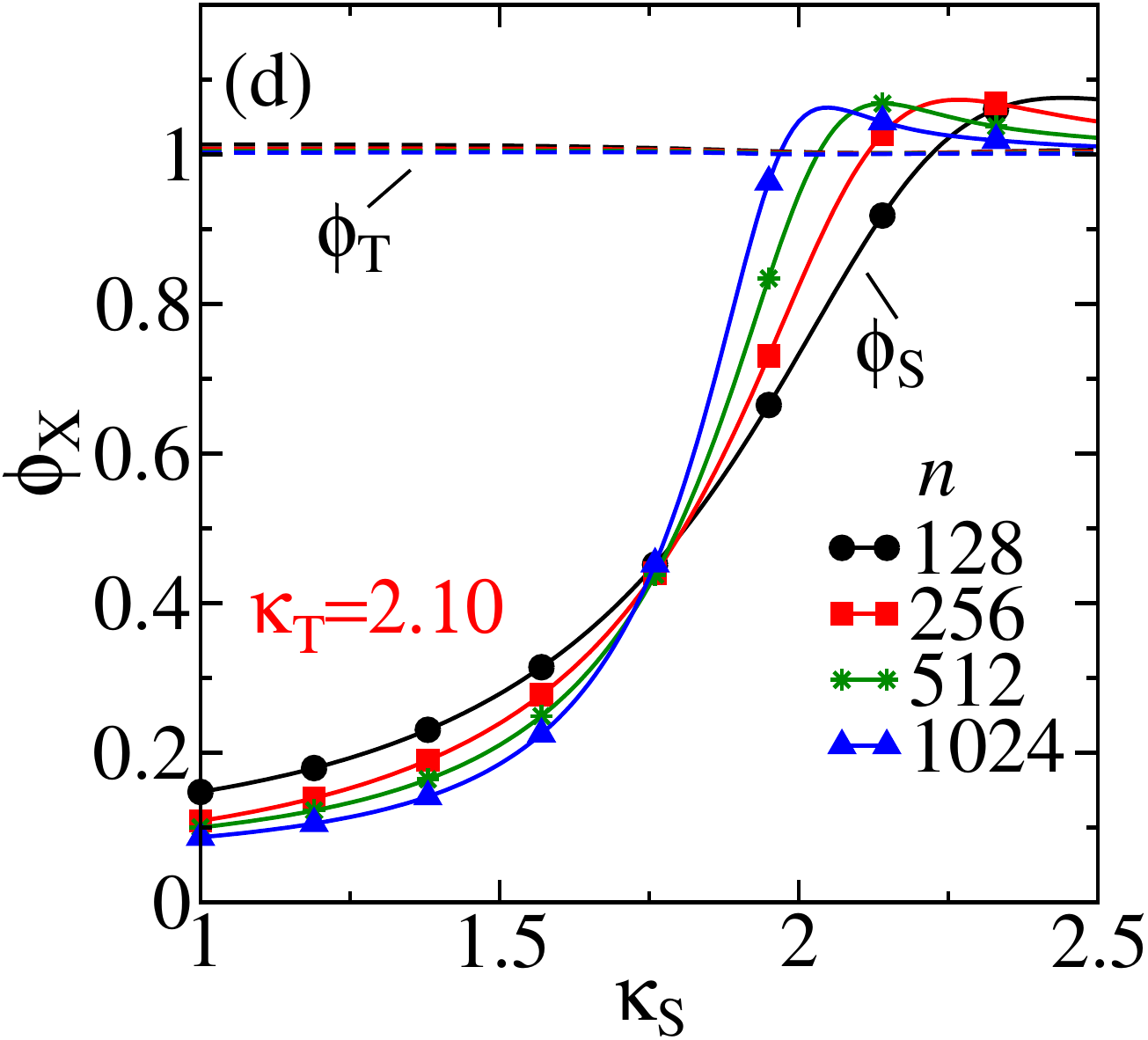}
    \caption{Results for vicinal surfaces with broad terraces and kinked steps featuring a distance between kinks of $\mathcal{D}=a=1$. Panels (a)--(c) present effective Flory exponents versus $\kappa_T$ ($\kappa_S$) for the indicated values of $\kappa_S$ ($\kappa_T$) kept fixed and a chain length of $n=1024$. Panel (d) displays the effective exponents $\phi_{S}$ (dashed lines) and $\phi_{T}$ (solid lines) versus $\kappa_S$, for $\kappa_T=2.10$ across several lengths $n$.}
    \label{figSM4}
\end{figure}

However, by considering the fraction of step contacts strictly relative to the total number of adsorbed monomers, as mandated by Eq.~\ref{EqUnalt}, the corresponding fluctuation metric $\Gamma_{S,n}$ is properly corrected to:
\begin{equation}
    \Gamma_{S,n}^* =  \frac{\langle m_S^2 \rangle-\langle m_S\rangle^2}{\langle m_S \rangle} - \frac{\langle m_S m_T \rangle-\langle m_S\rangle \langle m_T\rangle}{\langle m_T \rangle}.
    \label{EqGammaCorr}
\end{equation}
Consequently, it is reasonable to expect that the maximum of this adjusted function scales as:
\begin{equation}
    \Gamma_{S,n}^{*(max)} \sim n_a^{1/\delta_{S}}.
    \label{EqGammaMaxStep}
\end{equation}
As anticipated, the crossover exponents derived via this corrected scaling are far closer to the expected $1/2$. Nevertheless, in the immediate vicinity of the multicritical point, they still present slightly suppressed values --- for instance, one obtains $1/\delta_{S} \approx 0.3$ at $\kappa_T=1.4$ --- which implies that further, higher-order scaling corrections remain relevant in this specific regime. A substantially less plausible physical scenario would be that the exponent $1/\delta_{S}$ spontaneously deviates from $\phi_{S}$ solely in the neighborhood of the multicritical point.


\section{Results for vicinal surfaces with wide terraces and kinked steps}

Let us now consider vicinal surfaces where kinks are present along the steps, again assuming broad terraces. We have verified that the results for $1 < \mathcal{D} < n$ smoothly interpolate between those discussed above for straight steps ($\mathcal{D} > n$) and those for the limiting case of $\mathcal{D} = a = 1$. Therefore, we will concentrate exclusively on this latter case here, bearing in mind that analogous behaviors are observed for larger values of $\mathcal{D}$.

\begin{figure}[t]
    \centering
    \includegraphics[width=3.8cm]{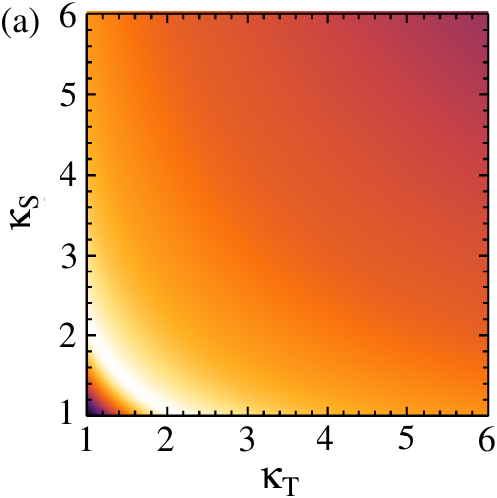}
    \includegraphics[width=4cm]{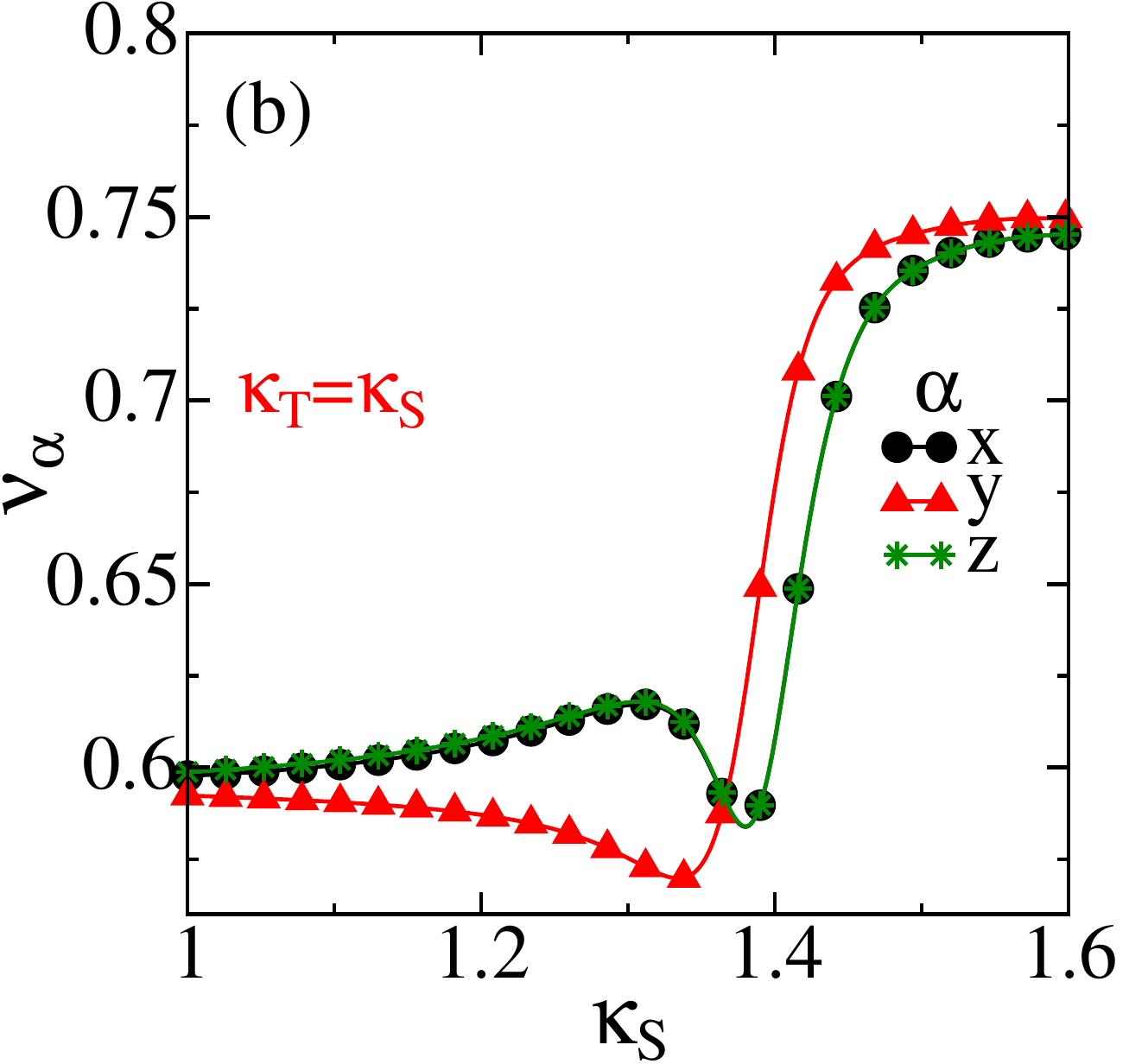}
    \includegraphics[width=4cm]{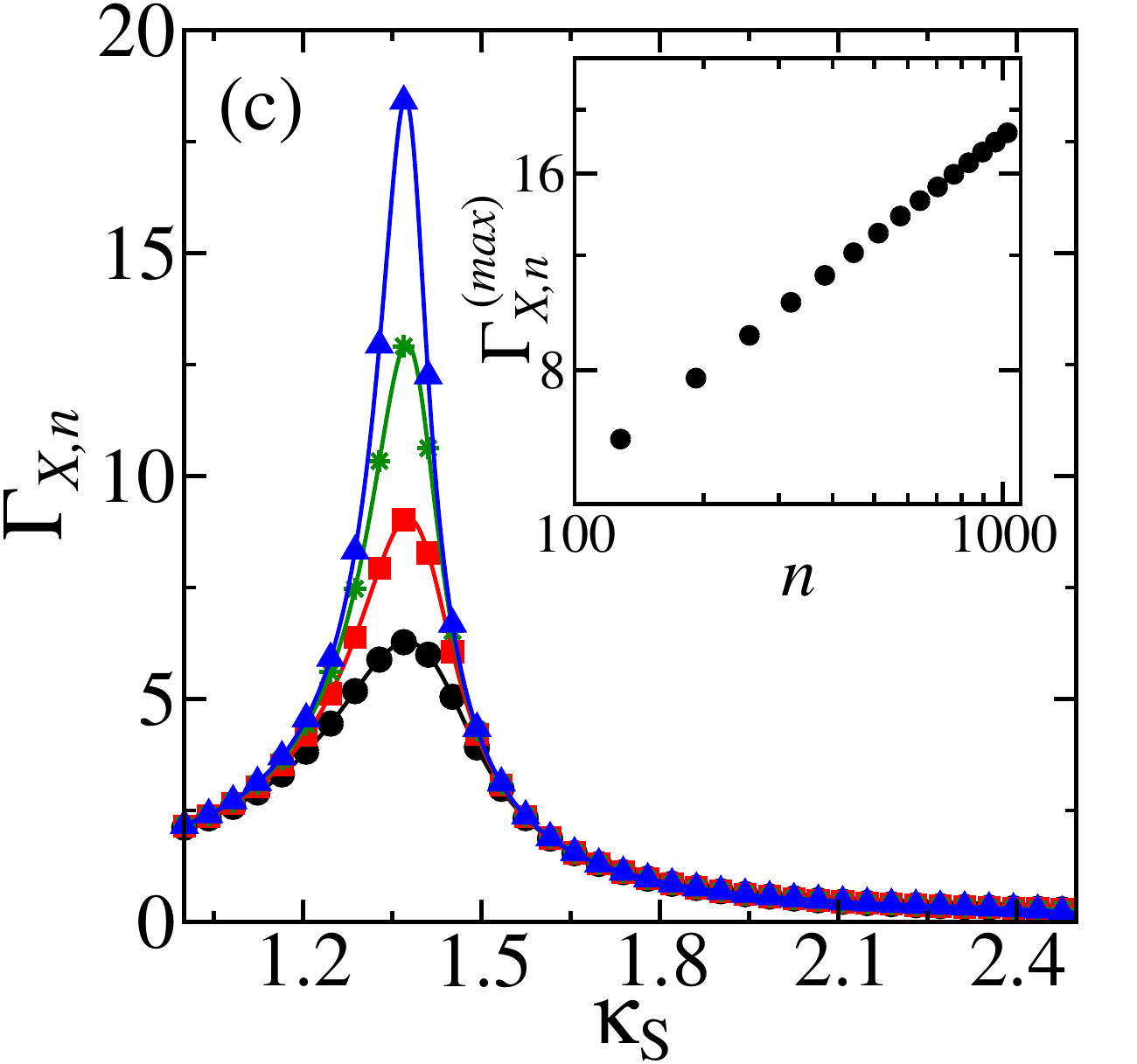}
    \includegraphics[width=4cm]{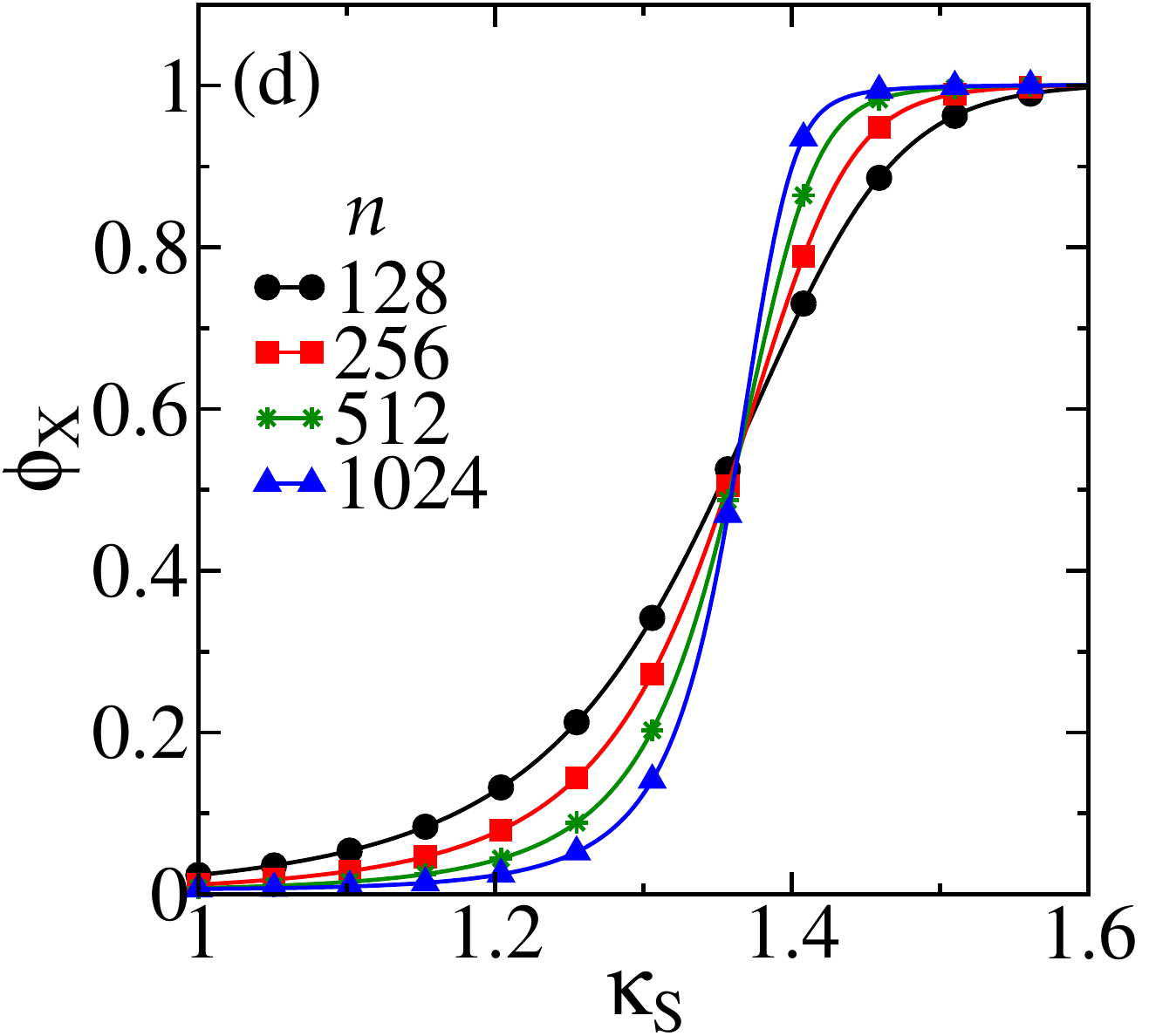}
    \caption{Results for vicinal surfaces with straight steps and a terrace width of $\mathcal{W}=a=1$. (a) Fluctuation map in the $\kappa_T$--$\kappa_S$ plane for $n=128$. Panels (b), (c), and (d) respectively display the effective Flory exponents (for $n=1024$), the fluctuations $\Gamma_{X,n}$, and the effective exponents $\phi_X$ (with $X=T,S$) as a function of $\kappa_S$, evaluated along the symmetric line $\kappa_T=\kappa_S$. No appreciable difference is observed between $\phi_T$ and $\phi_S$ in panel (d). The lengths in (c) match those indicated by the legend in (d).}
    \label{figSM5}
\end{figure}

Figure~\ref{figSM4} shows the Flory exponents at fixed $\kappa_T$ (or $\kappa_S$) for $\mathcal{D}=1$. As defined in the main text, our coordinate system is established such that the $x$- and $z$-axes are always perpendicular to the macroscopic step direction, while the $y$-axis remains parallel to it. Consequently, the scaling behavior closely mirrors the straight-step scenario [Figs.~\ref{figSM4}(a)--(c)]. Specifically, in the step-attached phase, the chain becomes highly extended along the step ($\nu_y \to 1$) while remaining strictly confined in the transverse direction ($\nu_x \to 0$). Furthermore, the surface exponents $\phi_{S,T}$ [Fig.~\ref{figSM4}(d)] and the crossover exponents $1/\delta_{S,T}$ (not shown) display the exact same scaling as the straight-step case, confirming that the underlying thermodynamics are functionally identical for both step geometries.


\begin{figure*}[t]
    \centering
    \includegraphics[width=3.5cm]{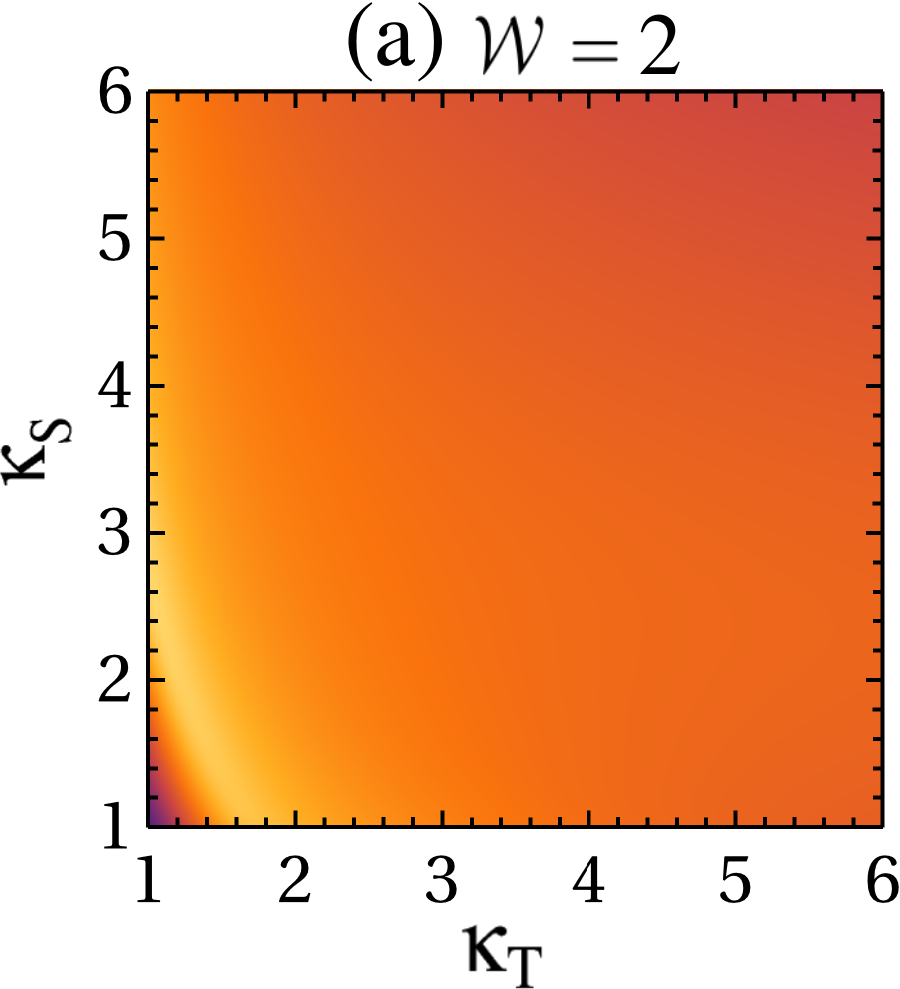}
    \includegraphics[width=3.5cm]{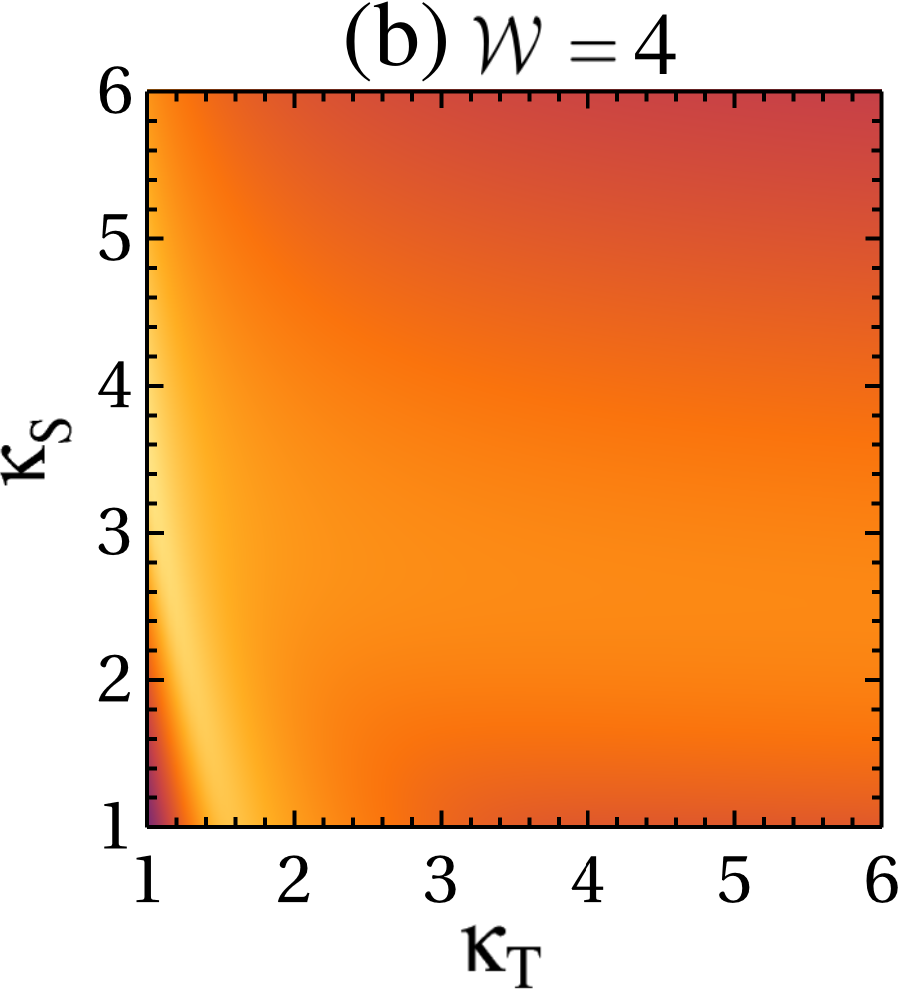}
    \includegraphics[width=3.5cm]{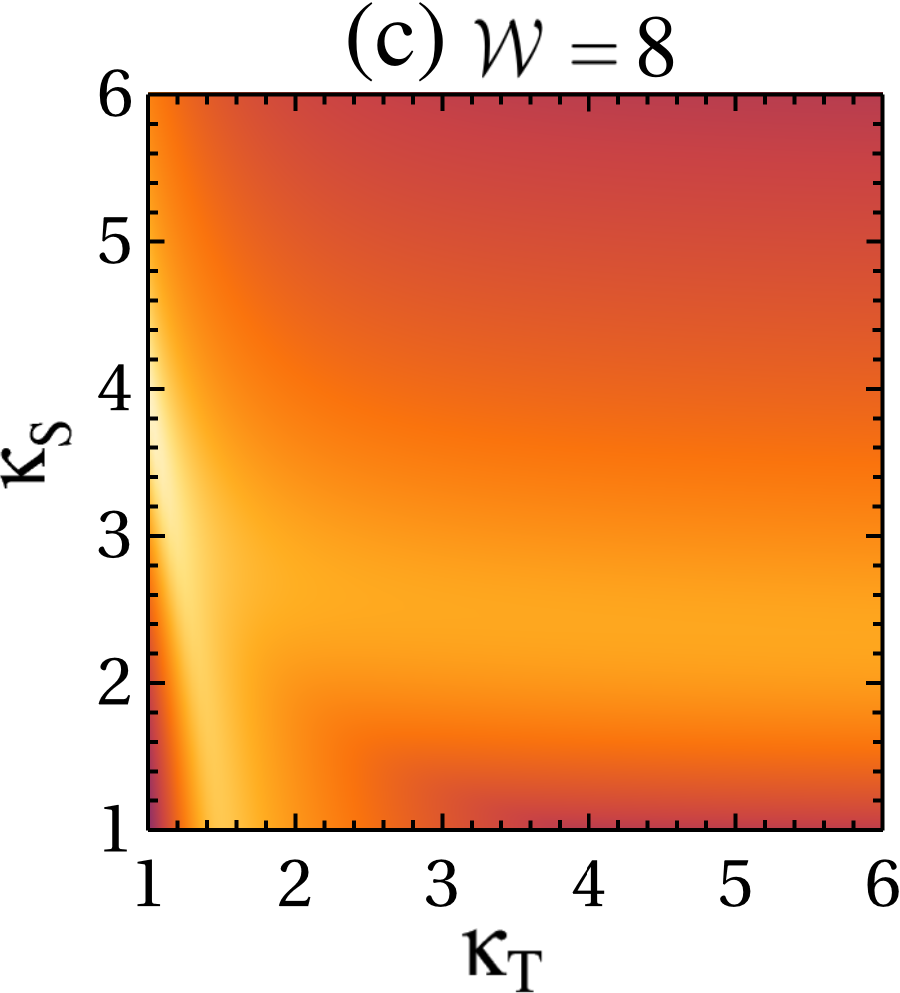}
    \includegraphics[width=3.5cm]{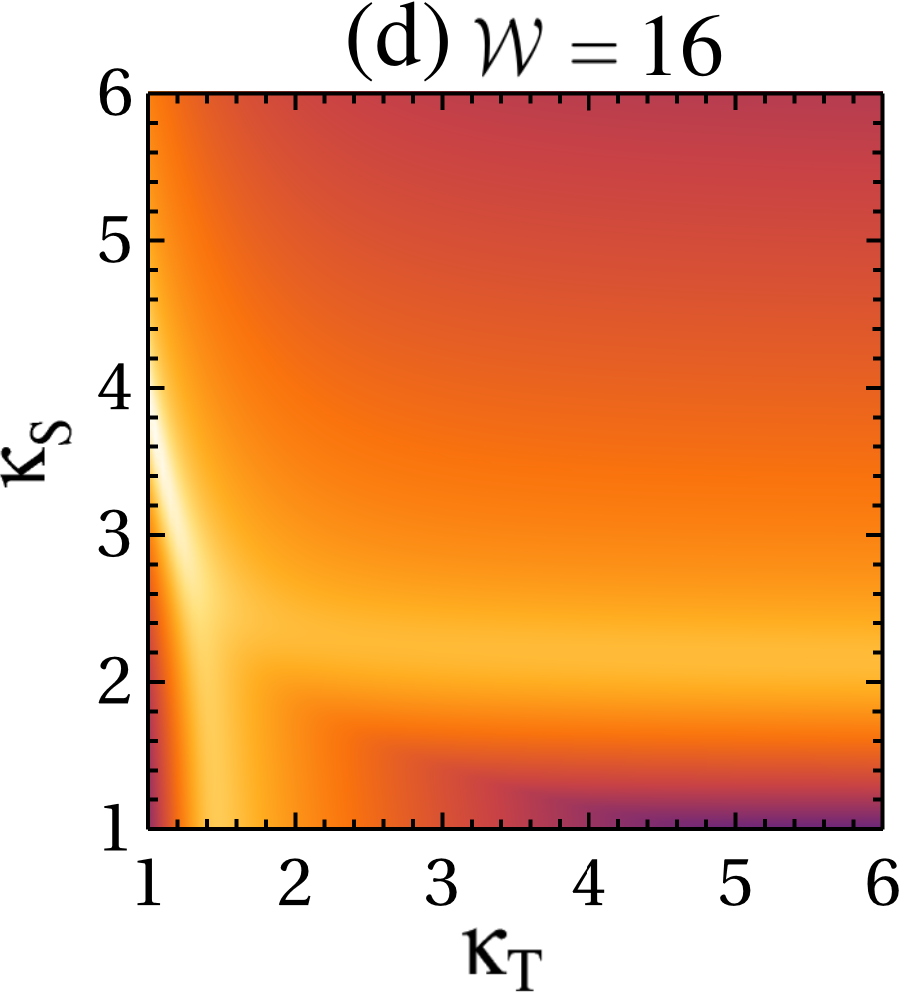}
    \includegraphics[width=3.5cm]{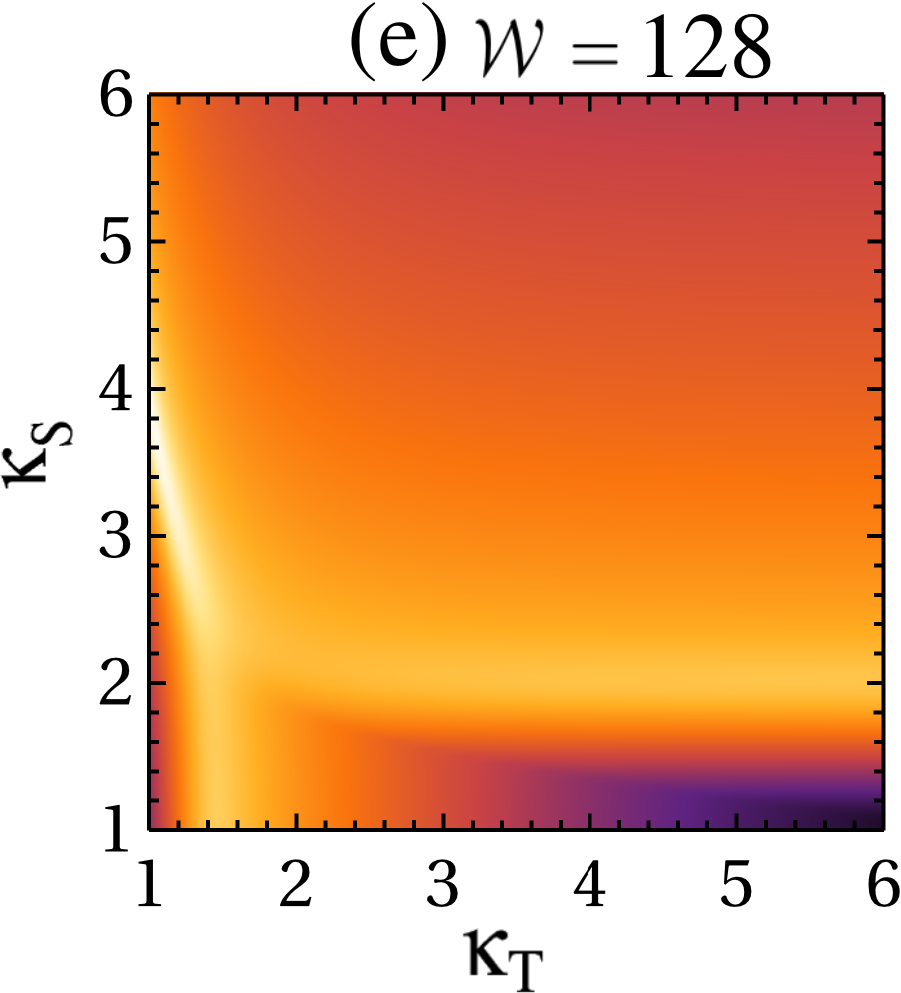}
    \caption{Results for vicinal surfaces with straight steps and short terraces. Fluctuation maps in the $\kappa_T$--$\kappa_S$ plane for a polymer length $n = 128$ and terrace widths of (a) $\mathcal{W} =2$, (b) $\mathcal{W} = 4$, (c) $\mathcal{W} = 8$, (d) $\mathcal{W} = 16$, and (e) $\mathcal{W} = 128$.}
    \label{figSM6}
\end{figure*}

\section{Results for vicinal surfaces with short terraces and straight steps}

While the wide-terrace situations analyzed in previous sections ensure that the polymer interacts with only a single step, in practice, a polymer may possess an average radius of gyration $R_g^{(2D)}$ large enough to simultaneously visit multiple steps. Hence, it is important to further investigate the thermodynamic properties of adsorption on substrates with very short ($\mathcal{W} = a = 1$) and intermediate ($2 \leq \mathcal{W} \lesssim R_g^{(2D)}$) terrace widths.

\subsection{The limiting case $\mathcal{W}=a=1$}

As discussed in the main text, steps and terraces become structurally equivalent when $\mathcal{W}=\mathcal{H}=a=1$; thus, the step-attached and terrace-adsorbed phases becomes a single adsorbed phase. Since each surface monomer carries an identical weight $\kappa = \kappa_S\kappa_T$, the desorbed-to-adsorbed transition strictly occurs at $\kappa_{S,c} = b / \kappa_{T,c}$, where $b = \kappa_c$. This symmetric desorbed-to-adsorbed boundary is clearly corroborated by the fluctuation map displayed in Fig.~\ref{figSM5}(a).

Figure~\ref{figSM5}(b) illustrates the variation of the Flory exponents with respect to $\kappa_S$, evaluated along $\kappa_T = \kappa_S$. Notably, instead of exhibiting $\nu_z=0$ in the adsorbed phase, one observes $\nu_z=\nu_x$ across both phases. This apparent anomaly arises because our coordinate system maintains the $z$-direction perpendicular to the individual microscopic terrace, while the global geometry of the $\mathcal{W}=1$ vicinal surface constitutes a 45$^\circ$ sloped staircase. Consequently, the local $z$-direction is no longer perpendicular to the macroscopic surface plane, rendering it functionally equivalent to the $x$-direction. Despite this geometric rotation, once the chain adsorbs onto the tilted plane, its spatial extension projects symmetrically across all relevant directions. Indeed, we find $\nu_x=\nu_z\approx \nu_y\approx \nu_{3D}$ for small $\kappa_S$, and $\nu_x=\nu_z\approx \nu_y \approx \nu_{2D}$ for large $\kappa_S$, exactly as expected for the desorbed and 2D-adsorbed phases, respectively. The crossing of the exponent curves precisely localizes the critical point at $\kappa_{S,c} = 1.367(2)$.

This finding is further corroborated by the behavior of the fluctuations $\Gamma_{X,n}$ [Fig.~\ref{figSM5}(c)] and the effective exponents $\phi_T$ and $\phi_S$ [Fig.~\ref{figSM5}(d)]. Therefore, since we are considering $\kappa_T = \kappa_S$, the global critical weight is determined to be $b = \kappa_c = \kappa_{S,c}^2 = 1.869(5)$. Moreover, we obtain critical exponents $\phi_{X,c}$ and $1/\delta_{X,c}$ (for $X = T, S$) that are in excellent agreement with $\phi_{c} = 1/\delta_{c} = 0.48$, as expected for 3D polymers adsorbing onto a flat 2D surface.

\begin{figure}[b]
    \centering
    \includegraphics[width=4cm]{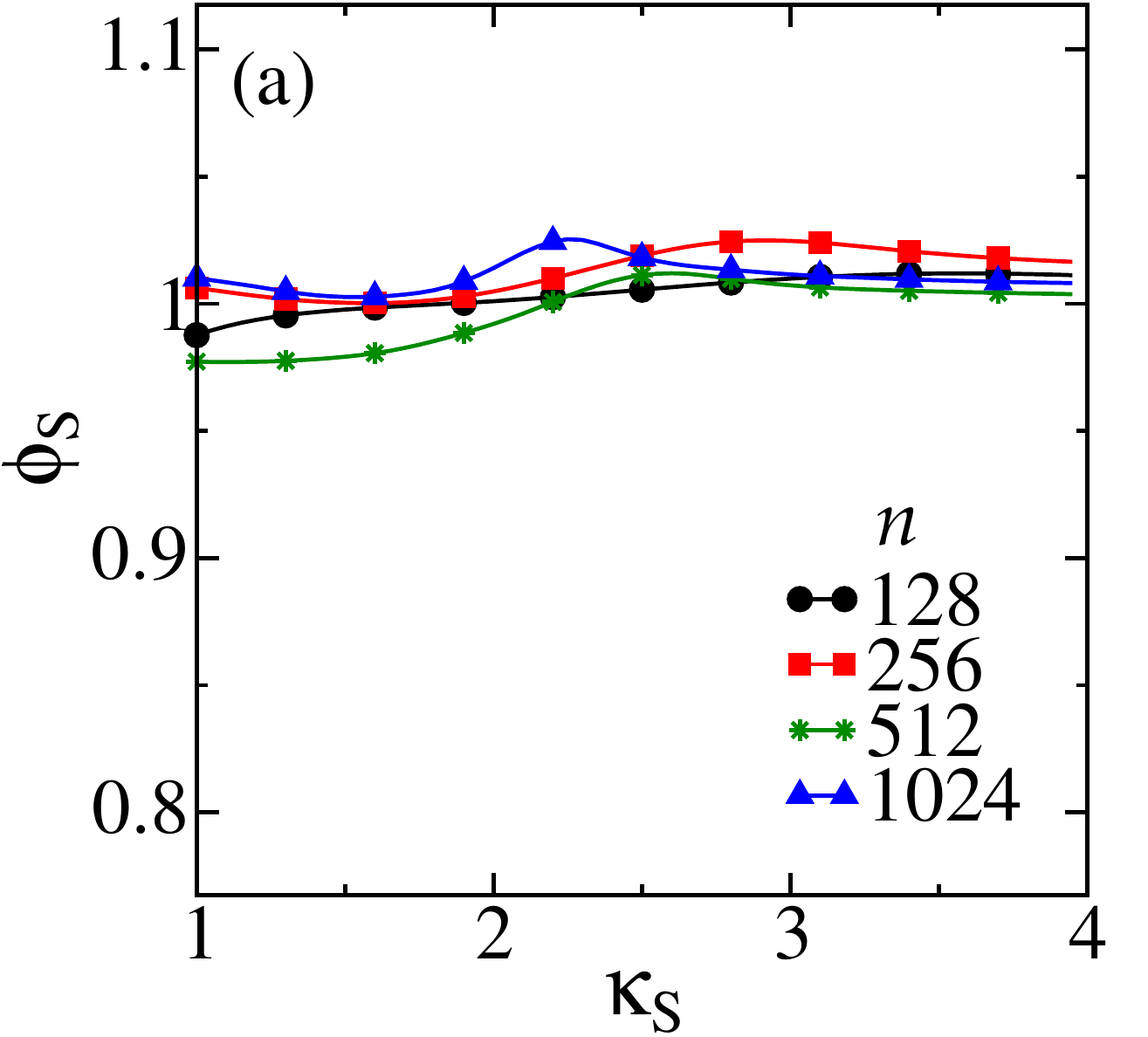}
    \includegraphics[width=4cm]{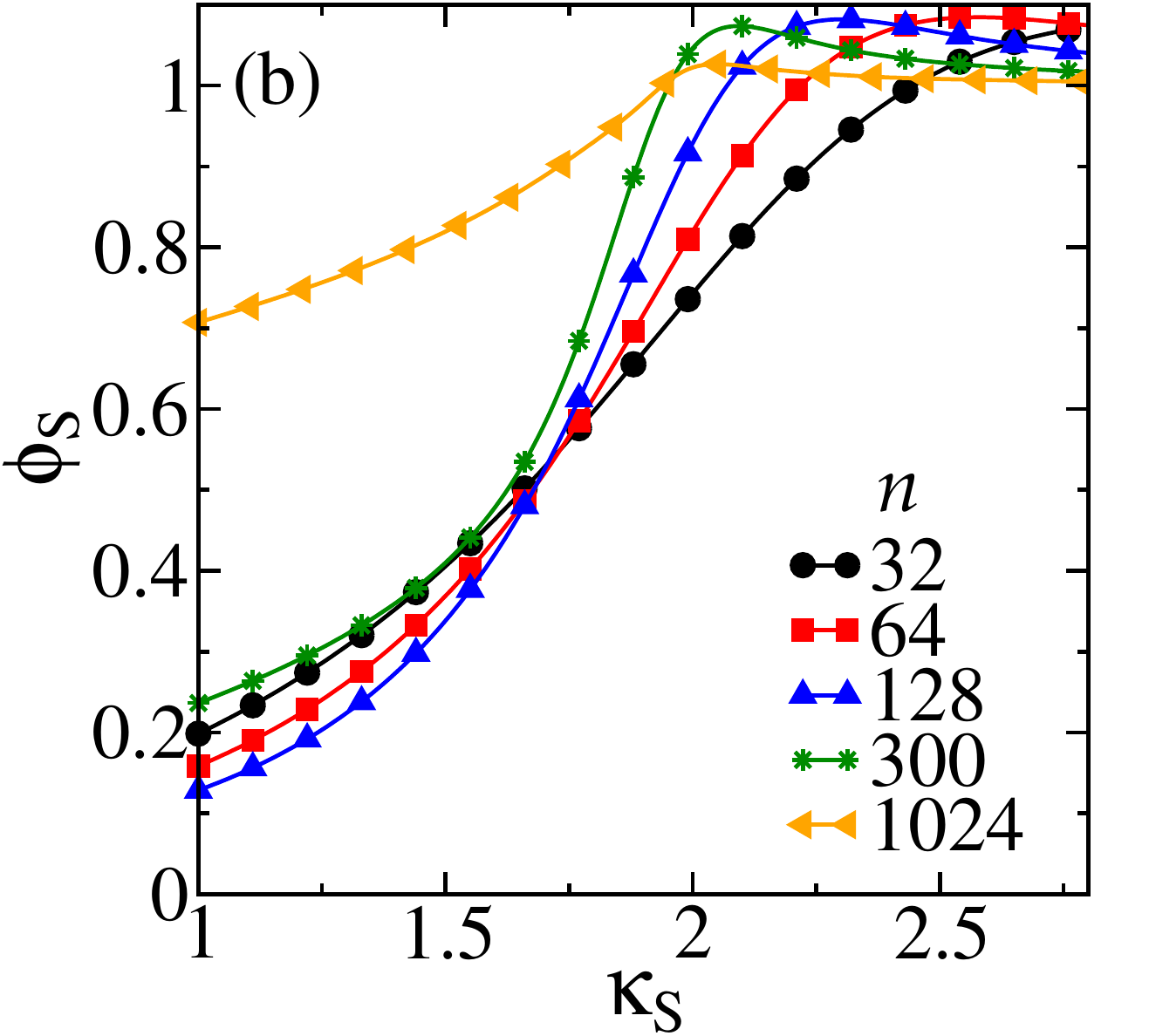}
    \includegraphics[width=4cm]{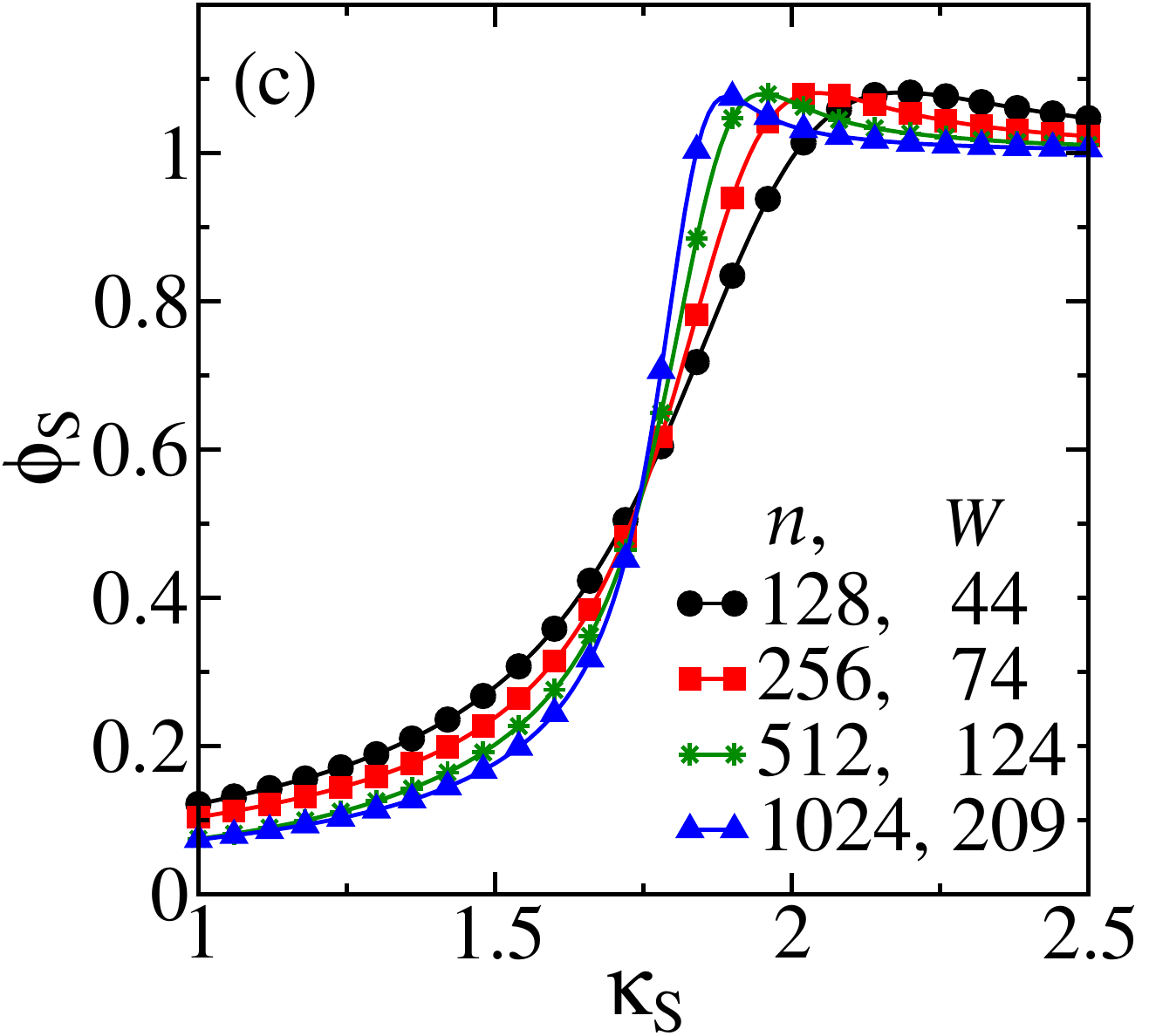}
    \includegraphics[width=4cm]{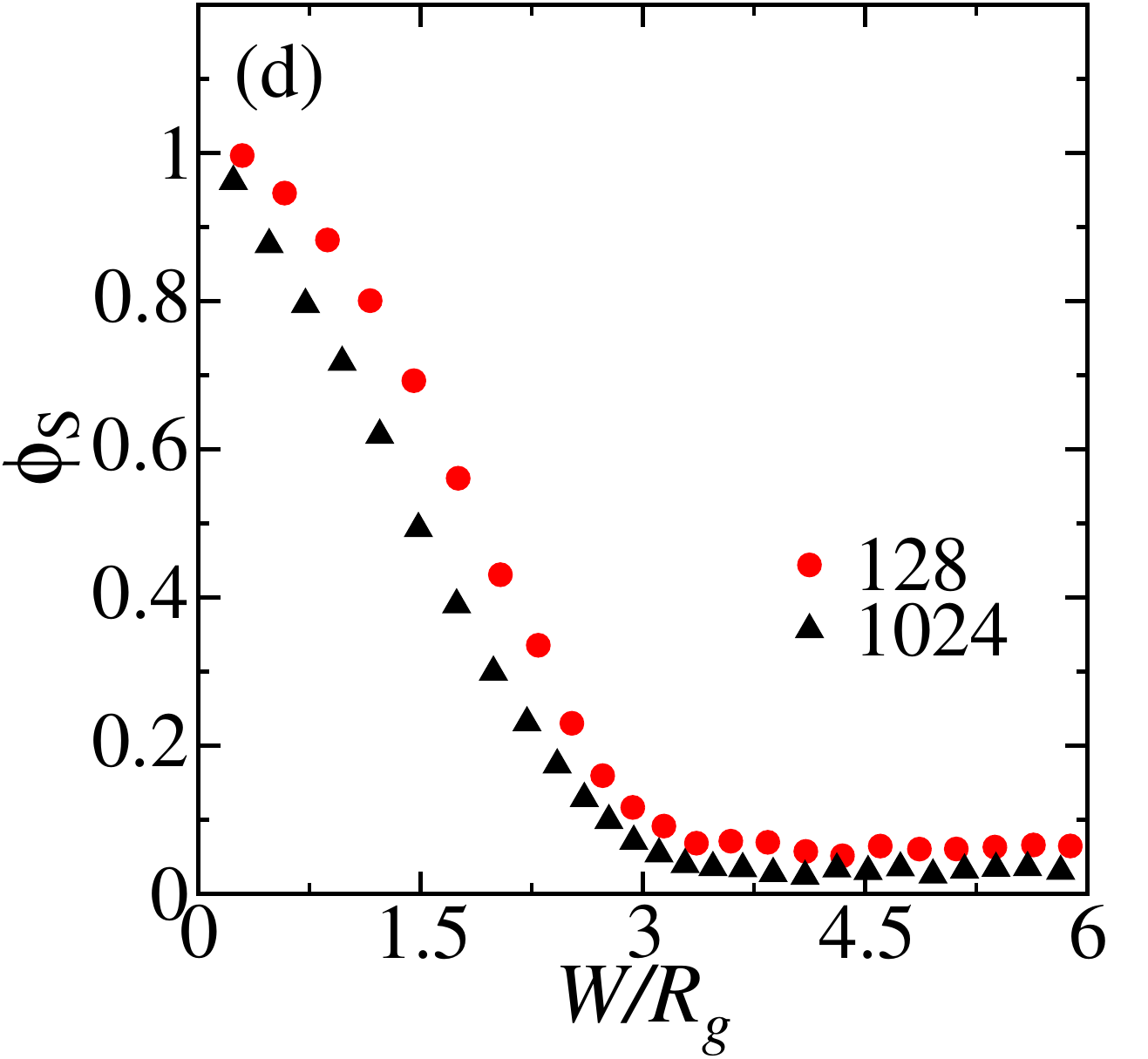}
    \caption{Results for vicinal surfaces with straight steps and short terraces. Panels (a)--(c) plot the effective exponent $\phi_S$ versus $\kappa_S$ for: (a) $\kappa_T = 2.0$ and $\mathcal{W}=19$; (b) $\kappa_T = 2.0$ and $\mathcal{W}=64$; and (c) $\kappa_T = 4.0$ utilizing varying widths $\mathcal{W}$ specifically tuned to maintain $\mathcal{W}/R_g^{(2D)} \approx 3.5$. Panel (d) demonstrates the variation of $\phi_S$ against the ratio $\mathcal{W}/R_g^{(2D)}$, evaluated at fixed parameters $\kappa_T=4.0$ and $\kappa_S=1.0$. The polymer lengths $n$ are detailed in the respective legends.}
    \label{figSM7}
\end{figure}

\subsection{Intermediate widths}

As indicated by the Flory exponents detailed in the main text, intermediate $\mathcal{W}$'s fundamentally suppress a phase transition between the terrace-adsorbed and step-attached phases, replacing it with a crossover. The fluctuation maps presented in Fig.~\ref{figSM6} for $n=128$ at increasing values of $\mathcal{W}$ clearly illustrate that only mild fluctuations occur at the boundary between these phases (visible as the faint, horizontal bright stripe separating the large and small $\kappa_S$ regions) when $\mathcal{W}$ is small. Notably, this boundary feature disappears entirely for $\mathcal{W} < 4$. Conversely, as $\mathcal{W}$ increases, the boundary fluctuations intensify and the crossover region narrows. From $\mathcal{W} = 16$ onward --- corresponding to widths larger than the average radius of gyration $R_g^{(2D)}$ --- no appreciable differences are discernible between the maps [see Figs.~\ref{figSM6}(d) and (e)]. This strongly implies that, for a fixed chain length $n$, increasing the terrace width eventually forces the crossover to sharpen into a genuine phase transition. Alternatively, by holding $\mathcal{W}$ fixed and increasing $n$ toward the thermodynamic limit, the system will asymptotically exhibit crossover behavior, even if small-$n$ data indicates the existence of a continuous transition.

Figure~\ref{figSM7} presents a series of effective exponent $\phi_S$ measurements that confirm this geometric constraint. For relatively narrow widths (e.g., $\mathcal{W} = 19$) and chain lengths $n \geq 128$ (yielding $\mathcal{W}/R_g^{(2D)} \lesssim 1.5$), we observe $\phi_S \approx 1$ deep inside both the terrace-adsorbed and step-attached regimes [Fig.~\ref{figSM7}(a)], verifying the absence of critical scaling as the system traverses the apparent phase boundary. Figure~\ref{figSM7}(b) plots $\phi_S$ versus $\kappa_S$ for a wider terrace ($\mathcal{W} = 64$); interestingly, the curves for $n \leq 128$ (where $\mathcal{W}/R_g^{(2D)} \gtrsim 5$) display intersection points characteristic of a critical transition, clustered near $\phi_{S,c} = 1/2$. However, the curves for $n=300$ and $1024$ (where the spatial constraint tightens to $\mathcal{W}/R_g^{(2D)} \approx 2.7$ and $\approx 1$, respectively) fail to cross the others, confirming that a crossover mechanism is recovered at sufficiently small $\mathcal{W}/R_g^{(2D)}$ ratios.

Conversely, if the length $n$ is systematically increased while holding the spatial ratio fixed at $\mathcal{W}/R_g^{(2D)} \approx 3.5$, for example, the hallmarks of criticality persist across all simulated sizes [Fig.~\ref{figSM7}(c)]. This provides robust evidence that a true continuous phase transition separates the terrace-adsorbed and step-attached states exclusively when $\mathcal{W}/R_g^{(2D)}$ is large, changing to a crossover when this ratio is small.

\begin{figure}[t]
    \centering
    \includegraphics[width=4.2cm]{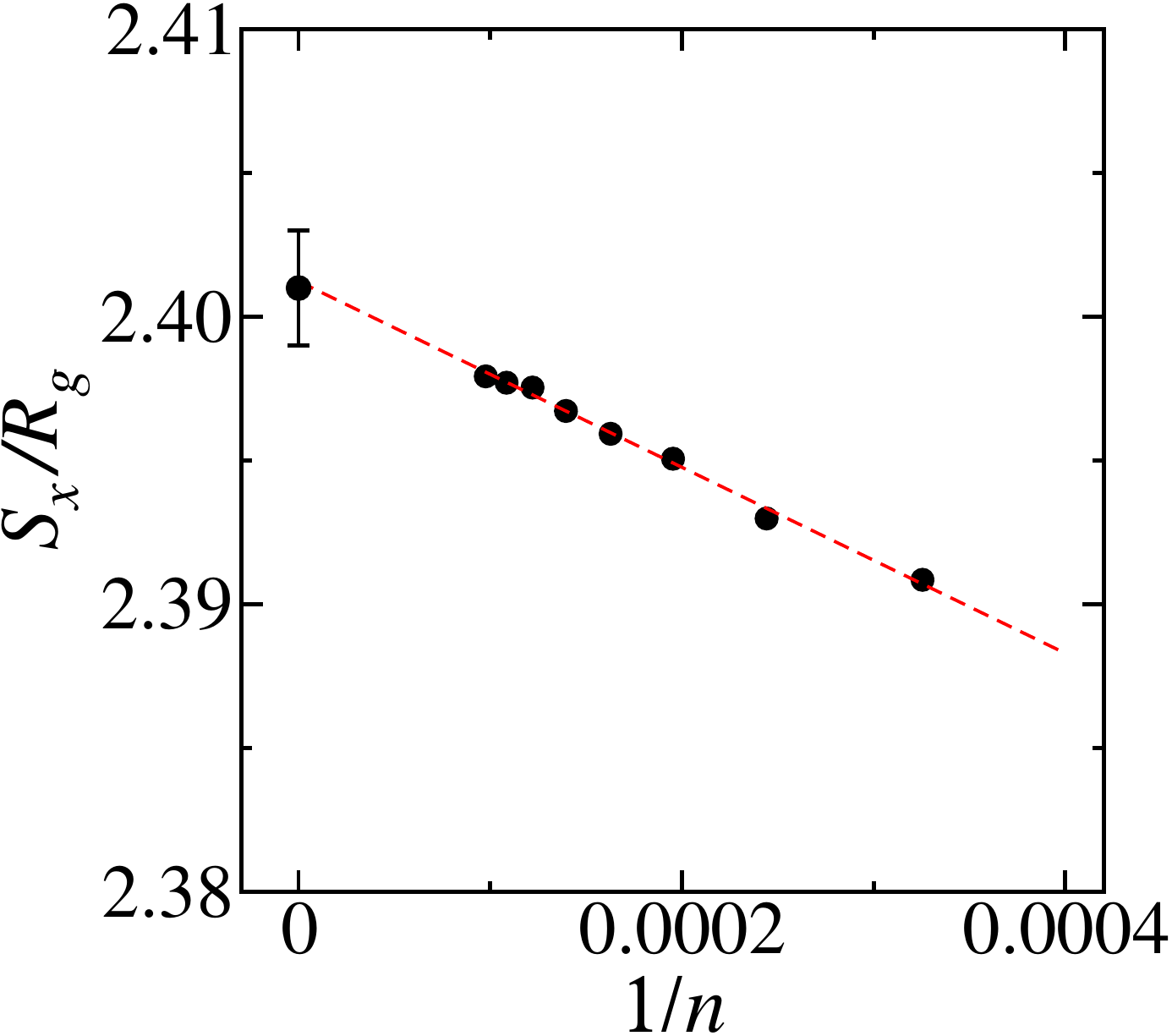}
    \caption{Extrapolation of the ratio between the mean transverse span and the radius of gyration, $S_x/R_g$, plotted as a function of $1/n$ for SAWs on a 2D square lattice.}
    \label{figSM8}
\end{figure}

Therefore, it is relevant to inquire: what threshold ratio $\mathcal{W}/R_g^{(2D)}$ demarcates these two distinct thermodynamic regimes? To answer this, we calculated the exponent $\phi_S$ deep inside the terrace-adsorbed phase for several $\mathcal{W}/R_g^{(2D)}$, fixing $n$ while tuning $\mathcal{W}$. As shown in Fig.~\ref{figSM7}(d), the effective exponents stop depending on the ratio and get close to zero only when $\mathcal{W}/R_g^{(2D)} \gtrsim 3$. Thus, the terrace-adsorbed phase is genuinely devoid of a macroscopic number of monomer-step contacts only at these large ratios. When $\mathcal{W}/R_g^{(2D)} \lesssim 3$, the polymer's extensive footprint makes it to traverse multiple steps, intrinsically imparting characteristics of the step-attached phase to the terrace-adsorbed state. Consequently, a sharp phase transition can only happen when $\mathcal{W}/R_g^{(2D)} \gtrsim 3$. This constraint is consistent with the transition observed at $\mathcal{W}/R_g^{(2D)} \approx 3.5$ [Fig.~\ref{figSM7}(c)] and the crossover seen at $\mathcal{W}/R_g^{(2D)} \approx 2.7$ [Fig.~\ref{figSM7}(b)].

Since the concurrent crossing of multiple steps physically prevents the transition, we are naturally led to ask: how many steps must a polymer interact with before the critical behavior turns into a crossover? Equivalently, how many individual terraces does the chain span when $\mathcal{W} \approx 3 R_g^{(2D)}$? While $R_g^{(2D)}$ effectively gauges the root-mean-square spatial extent, the dimension that strictly governs step-crossing is the transverse mean span (caliper size) $S_x$, which measures the absolute mean distance between the polymer's geometrical extremities along the $x$-direction. Although $S_x$ is known to share the same scaling as $R_g^{(2D)}$ (i.e., $S_{x} \simeq B n^{\nu_{2D}}$ vs. $R_g^{(2D)} \simeq A n^{\nu_{2D}}$), the universal amplitude ratio $S_{x}/R_g^{(2D)} \simeq B/A$ for 2D SAWs on the square lattice has, to our knowledge, never been quantified. To resolve this, we simulated SAWs on a square lattice for chain lengths up to $n=10240$, measuring both $S_x(n)$ and $R_g(n)$, and their ratio is presented in Fig.~\ref{figSM8} as a function of $1/n$. This specific choice of scaling variable effectively linearizes the data, enabling a extrapolation to the thermodynamic limit ($n \to \infty$) which yields $S_x/R_g^{(2D)} = 2.401(2)$. This reveals that the geometric condition required to sustain a continuous phase transition between the two adsorbed phases is approximately $\mathcal{W} \gtrsim 1.2 S_x$. In physical terms, the polymer strictly cannot span more than a single terrace for the continuous transition to occur.